\documentclass[a4paper,UKenglish,cleveref, autoref, thm-restate,final]{lipics-v2021}
\hideLIPIcs  %uncomment to remove references to LIPIcs series (logo, DOI, ...), e.g. when preparing a pre-final version to be uploaded to arXiv or another public repository

\title{Permissive Equilibria in Multiplayer Reachability Games} %TODO Please add

\author{Aline Goeminne}{F.R.S.-FNRS \& UMONS -- Université de Mons, Belgium }{aline.goeminne@umons.ac.be}{}{Postdoctoral Researcher of the Fonds de la Recherche Scientifique -- FNRS.}%TODO mandatory, please use full name; only 1 author per \author macro; first two parameters are mandatory, other parameters can be empty. Please provide at least the name of the affiliation and the country. The full address is optional. Use additional curly braces to indicate the correct name splitting when the last name consists of multiple name parts.

 \author{Benjamin Monmege}{Aix-Marseille Univ, CNRS, LIS, Marseille, France}{benjamin.monmege@univ-amu.fr}{}{This author was partially funded by ANR JCJC Quasy ANR-23-CE48-0008}

\authorrunning{A. Goeminne and B. Monmege} %TODO mandatory. First: Use abbreviated first/middle names. Second (only in severe cases): Use first author plus 'et al.'

\Copyright{Aline Goeminne and Benjamin Monmege} %TODO mandatory, please use full first names. LIPIcs license is "CC-BY";  http://creativecommons.org/licenses/by/3.0/

 %TODO mandatory; please add comma-separated list of keywords

\category{} %optional, e.g. invited paper

\relatedversion{} %optional, e.g. full version hosted on arXiv, HAL, or other respository/website
\nolinenumbers %uncomment to disable line numbering

\EventEditors{John Q. Open and Joan R. Access}
\EventNoEds{2}
\EventLongTitle{42nd Conference on Very Important Topics (CVIT 2016)}
\EventShortTitle{CVIT 2016}
\EventAcronym{CVIT}
\EventYear{2016}
\EventDate{December 24--27, 2016}
\EventLocation{Little Whinging, United Kingdom}
\EventLogo{}
\SeriesVolume{42}
\ArticleNo{23}
\newtheorem{problem}{Problem}

\newcommand{\aline}[2][noinline]{\todo[linecolor=blue,backgroundcolor=white,bordercolor=blue, #1]{{\color{blue}AG:} #2}}

\newcommand{\IE}{\emph{i.e., }}
\newcommand{\N}{\mathbb{N}}

\DeclareMathOperator{\playerSet}{N}
\newcommand{\weight}{w}
\DeclareMathOperator{\plays}{Plays}
\DeclareMathOperator{\hist}{Hist}
\DeclareMathOperator{\last}{Last}
\DeclareMathOperator{\game}{\mathcal{G}}
\DeclareMathOperator{\gain}{Gain}
\newcommand{\initGame}{(\game,v_0)}
\DeclareMathOperator{\targetSet}{F}
\newcommand{\outcome}[2]{\langle  #1 \rangle_{#2}}
\DeclareMathOperator{\multiStrat}{\Theta}
\DeclareMathOperator{\multiActions}{A}
\newcommand{\initVertex}{v_0}

\newcommand{\multiOutcome}[2]{\outcome{#1}{#2}}
\newcommand{\multiHistory}[2]{\outcome{#1}{#2}^{\mathrm {H}}}
\DeclareMathOperator{\tree}{\mathcal{T}}
\DeclareMathOperator{\infTree}{\tree^\infty}
\DeclareMathOperator{\depth}{depth}
\DeclareMathOperator{\height}{height}
\DeclareMathOperator{\penalAux}{Penalty}
\newcommand{\penal}[3]{\penalAux_{#1}^{#2}(#3)}
\newcommand{\initSubgame}[2]{(\game_{\restriction #1},#2)}
\DeclareMathOperator{\successor}{Succ}

\DeclareMathOperator{\forest}{\mathcal{F}}

\DeclareMathOperator{\Penalty}{Penalty}
\DeclareMathOperator{\mainPenalty}{MPenalty}
\DeclareMathOperator{\retaliationPenalty}{RPenalty}

\DeclareMathOperator{\winningPlayers}{Win}

\newcommand{\initVisit}{\I_0}

\DeclareMathOperator{\visit}{Visit}
\DeclareMathOperator{\blocked}{Blocked}
\DeclareMathOperator{\deviators}{D}
\DeclareMathOperator{\I}{I}
\DeclareMathOperator{\J}{J}

\DeclareMathOperator{\Out}{Out}

\usepackage[dvipsnames]{xcolor}
\usepackage[obeyFinal]{todonotes}

\usepackage{tikz}
\usetikzlibrary{automata}
\usetikzlibrary{decorations.pathmorphing}
\tikzset{>=latex}

\begin{document}

\maketitle

\begin{abstract}
We study multi-strategies in multiplayer reachability games played on finite graphs. A multi-strategy prescribes a set of possible actions, instead of a single action as usual strategies: it represents a set of all strategies that are consistent with it. We aim for profiles of multi-strategies (a multi-strategy per player), where each profile of consistent strategies is a Nash equilibrium, or a subgame perfect equilibrium. The permissiveness of two multi-strategies can be compared 
with penalties, as already used in the two-player zero-sum setting by Bouyer, Duflot, Markey and Renault~\cite{BDMR09}. We show that we can decide the existence of a multi-strategy that is a Nash equilibrium or a subgame perfect equilibrium, while satisfying some upper-bound constraints on the penalties in PSPACE, if the upper-bound penalties are given in unary. The same holds when we search for multi-strategies where certain players are asked to win in at least one play or in all plays. 
\end{abstract}

%----
% Introduction
%----

\section{Introduction}

Nowadays, computer systems are ubiquitous and increasingly complex. Errors in such systems can have dramatic consequences. This is why \emph{model checking} provides a formal tool to ensure these systems are \emph{correct} and meet certain \emph{specifications}. \emph{Synthesis}, on the other hand, allows for the construction of a correct-by-construction system model: concepts from \emph{game theory} can be used for this purpose.

\emph{Two-player zero-sum games} are commonly used to model a system interacting with its environment. In this model, the system aims to achieve a goal while the environment acts antagonistically to prevent it. This situation can be abstracted as a \emph{game played on a graph} involving two players (the system and the environment). The graph represents the different possible configurations of the system, and an infinite path in this graph is a sequence of interactions between the system and the environment. In this model, building a correct system amounts to synthesizing a winning strategy, that is, a way for the system to play that ensures its goal is met regardless of the environment's behavior.

Unlike the purely antagonistic view of two-player zero-sum games, \emph{multiplayer games} allow for modeling situations where the environment may have its own goals, or where the system consists of different interacting components, each with its own specification. In this context, the notion of a winning strategy is no longer appropriate, hence notions of equilibria are studied: Nash equilibria or subgame perfect equilibria, which more adequately account for the sequential aspect of games played on graphs (avoiding non-credible threats). Intuitively, an equilibrium can be seen as a contract among players such that no player has an incentive to unilaterally change his strategy.

It is well known that different equilibria can coexist in the same game. In particular, a game may include an equilibrium where no player achieves his goal and an equilibrium where all players achieve their goals. The latter equilibrium is more \emph{relevant} than the former. Therefore, it seems appropriate to focus on the existence and synthesis of relevant equilibria (according to certain relevance criteria).

Even if the synthesis process provides an equilibrium, its implementation may fail. This can be due to the occurrence of errors; for example, the action prescribed by the equilibrium may be unavailable. Synthesizing \emph{robust equilibria} against such perturbations is therefore essential.
To address these robustness issues, the classic notion of a player’s strategy can be replaced by the notion of a \emph{multi-strategy}: unlike a classic strategy that provides a single action at each decision point, a multi-strategy provides a subset of possible actions (see, for example, \cite{BJW02,BDMR09}).

Intuitively, a multi-strategy is more \emph{permissive} than another if the first allows more behaviors than the second. There are different ways to express this permissiveness. A~qualitative view of permissiveness is studied in \cite{BJW02}, where a multi-strategy is more permissive than another if the set of resulting plays includes those of the second multi-strategy. A~quantitative view is addressed in \cite{BDMR09} via the notion of \emph{penalty} of multi-strategies, where a cost is associated with each edge not chosen by the multi-strategy. Thus, the penalty of a multi-strategy is the highest sum of blocked edges along a play consistent with the multi-strategy.

\subparagraph*{Related works} In \cite{BJW02}, permissiveness in parity games (a highly expressive winning condition) is studied: considering the qualitative view of permissiveness, there does not necessarily exist a most permissive strategy. However, one exists when restricted to memoryless strategies (which always make the same decision in any given vertex of the game). By reducing to safety games, the authors show that it is possible to compute the most permissive strategy. In \cite{BDMR09}, the above-mentioned quantitative view of permissiveness is implemented. Several penalty measures and games are used, and the complexity of computing the most permissive strategies in this context is given. More general parity objectives are then studied in \cite{BouyerMOU11}. Recently, other methods have explored permissiveness in two-player games using templates to concisely represent multiple strategies in graph games \cite{AnandNayakSchmuck}. This approach is also used in multiplayer games for the synthesis of secure equilibria \cite{NayakS24}.

Independently, different equilibria (Nash or subgame perfect) have been studied in multiplayer games to ensure a strategy profile where no player has an incentive to deviate. Several works have characterized such equilibria and studied the complexity of decision problems related to the existence of relevant equilibria. Notably, these works have focused on the study and characterization of ($i$) Nash equilibria in games where players have classical $\omega$-regular objectives \cite{Ummels06, BrihayeBGT19}, ($ii$) weak subgame perfect equilibria (a variant of subgame perfect equilibria) where players have classical $\omega$-regular objectives \cite{weakSPEs} (this work also characterizes subgame perfect equilibria when the studied objectives are either qualitative reachability or safety objectives); ($iii$) subgame perfect equilibria for games with quantitative reachability objectives \cite{BrihayeBGRB19}; ($iv$) subgame perfect equilibria for games with parity objectives~\cite{brice2022} or ($v$)~mean-payoff objectives \cite{Brice,briceMeanPayoff}. 

\subparagraph*{Contribution} Our goal is to combine these two research directions by studying permissiveness in strategy profiles that describe equilibria (Nash or subgame perfect). In this first work, we focus on reachability games only. We study permissive strategy profiles such that all the fully described strategy profiles they contain are equilibria. The motivation is to allow greater latitude and robustness of equilibrium profiles without losing quality in the final goal of secure synthesis. With the qualitative view, as in the two-player game framework \cite{BJW02}, it is not difficult to show that there does not necessarily exist most permissive profiles that are Nash equilibria (or subgame perfect equilibria). We will thus consider a quantitative view of permissiveness similar to the penalty measures introduced in \cite{BDMR09} for two-player games. We obtain a characterization based on trees, and decision algorithms with penalties bounded by a given threshold in polynomial space with respect to the size of the game and the maximal penalty bound (if this is encoded in unary). We also solve the problem of synthesis of robust and relevant equilibria, where the relevance is the constraint that all derived equilibria ensure that all players in a fixed subset satisfy their objective (\emph{strongly winning}), or that at least one derived equilibrium ensures this guarantee (\emph{weakly winning}).

All missing proofs can be found in the appendix.

%---
% Preliminaries
%----

\section{Multiplayer reachability games}\label{sec:prelim}

A \emph{(multiplayer) reachability games} is a tuple 
$(\playerSet, V, (V_i)_{i\in \playerSet}, E, (\targetSet_i)_{i\in \playerSet}, v_0)$, that we denote $\initGame$ to emphasize the $v_0$ component, where $\playerSet = \{ 1, \ldots, n\}$ is a finite set of $n$ players, $(V,E)$ is a finite directed graph without deadlocks  (for all $v \in V$, there exists $v' \in V$ such that $(v,v') \in E$), $(V_i)_{i \in \playerSet}$ is a partition of $V$ between the players,  $\targetSet_i \subseteq V$ is the set of target vertices, called \emph{target set}, of player $i \in \playerSet$, and $\initVertex$ is an initial vertex. Given a vertex $v \in V$, we let $\successor(v) = \{ v' \in V \mid (v,v') \in E \}$ be the set of all successors of $v$. 

A \emph{play} in $\game$ is an infinite sequence of vertices consistent with the graph structure, \IE if $\rho = \rho_0\rho_1 \cdots$ is a play, then for all $k \in \N$, $\rho_{k} \in V$ and $(\rho_k,\rho_{k+1}) \in E$. The set of plays is denoted by $\plays$, while $\plays(v)$ denotes the set of plays beginning in $v$.
Given a play $\rho=\rho_0\rho_1\cdots$ and $k \in \N$, $\rho_{\geq k}$ is the suffix $\rho_k\rho_{k+1}\cdots$ of $\rho$.

For each player $i \in \playerSet$, we let $\gain_i$ be the \emph{gain function} that associates with each play the value $1$ if the play is winning for player~$i$, $0$ if it is losing. For a reachability game as above, we have $\gain_i(\rho) = 1$ iff player $i$ reaches his target set in $\rho$, \IE $\rho = \rho_0\rho_1 \cdots$ and there exists $k \in \N$ with $\rho_k \in \targetSet_i$. \textbf{In the rest of this article, $\initGame$ will always denote a reachability game associated with these gain functions.}

A \emph{history} is a finite sequence of vertices $h= h_0h_1\cdots h_k$ with $k \in \N$ defined similarly.
The set of histories is denoted by $\hist$, while $\hist(v)$ denotes the set of histories beginning in~$v$.
For all $i \in \playerSet$, we write $\hist_i$ to denote the set of histories ending in a vertex owned by player~$i$. 
If $h = h_0\cdots h_k$ with $k \in \N$ is a history, $\last(h)$ denotes the last vertex $h_k$, while $|h|$ denotes its length $k$.
Given a history $h = h_0\cdots h_k$, $\visit(h) = \{ i \in \playerSet \mid \exists 1 \leq \ell \leq k \;  h_\ell \in F_i \}$ is the set of players who visit their target set along $h$.

A \emph{strategy} of player $i$ is a function $\sigma_i\colon \hist_i(\initVertex) \rightarrow V$ that assigns to each history $hv \in \hist_i(\initVertex)$ a vertex $v'$ such that $(v,v') \in E$. A play $\rho = \rho_0\rho_1\cdots$ is consistent with a strategy $\sigma_i$ if for all $\rho_k \in V_i$, $\sigma_i(\rho_0\cdots \rho_k) = \rho_{k+1}$. A \emph{strategy profile} is a tuple $\sigma = (\sigma_i)_{i\in \playerSet}$ of strategies, one per player: there is a unique play from $\initVertex$ which is consistent with each strategy $\sigma_i$, and we call this play the \emph{outcome} of $\sigma$, 
denoted by~$\outcome{\sigma}{\initVertex}$.
To highlight the role of player~$i$, we sometimes write $\sigma = (\sigma_i,\sigma_{-i})$ where $\sigma_{-i}$ denotes the strategy profile of the players other than player~$i$.

The strategy profile $\sigma$ is a \emph{Nash equilibrium} (NE) in $\initGame$ if no player has an incentive to deviate unilaterally from his strategy to increase his gain, \IE  if for all players $i \in \playerSet$ and all strategies $\sigma'_i$ of player~$i$, $\gain_i(\outcome{\sigma}{v_0}) \geq \gain_i(\outcome{\sigma'_i,\sigma_{-i}}{v_0})$.

The concept of \emph{subgame perfect equilibrium} (SPE) takes more into account the sequential nature of games played on graphs by avoiding non-credible threat, a well-known weakness of NEs in this setting.
Informally, a strategy profile is an SPE if it is an NE in all subgames.
Given a history $hv \in \hist(\initVertex)$, the \emph{subgame} $\initSubgame{h}{v}$ is obtained from $\game$ by changing the initial vertex to $v$, and by considering the gain functions $(\gain_{i\restriction h})_{i\in \playerSet}$ taking into account the players that have won in history $h$: we thus write, for each $i \in \playerSet$, $\gain_{i\restriction h}(\rho) = \gain_i(h\rho)$ for all $\rho \in \plays(v)$.
Moreover, if $\sigma_i$ is a strategy of player $i$ in $\game$, then $\sigma_{i \restriction h}$ is the strategy of player $i$ in the subgame $\initSubgame{h}{v}$ such that for all $h' \in \hist_i(v)$, $\sigma_{i\restriction h}(h') = \sigma_i(hh')$. In the same way, from a strategy profile $\sigma$ in $\game$, we can derive a strategy profile $\sigma_{\restriction h}$ in $\initSubgame{h}{v}$.
We now define formally the concept of SPEs: 
a strategy profile $\sigma$ is an SPE in $\game$ if for all $i \in \playerSet$, for all $hv \in \hist_i(\initVertex)$, 
$\sigma_{\restriction h}$ is an NE in $\initSubgame{h}{v}$. Notice that an SPE is an NE and that there always exists an SPE (and thus an NE) in a reachability game~\cite{Ummels06}.

\section{Permissiveness in strategies}\label{sec:permissiveness}

Our goal is to allow for some permissiveness in strategies of all players, \IE being able to underspecify the strategies of the players, while maintaining that they describe an NE or an SPE.

A \emph{multi-strategy} of player~$i$ is a function $\multiStrat_i\colon \hist_i(\initVertex) \rightarrow 2^V \setminus \{ \emptyset \}$ that assigns to each history $hv \in \hist_i(\initVertex)$ a non-empty set of vertices $\multiActions \subseteq V$ such that for all $v' \in \multiActions$, $(v,v') \in E$. 
Notice that a strategy $\sigma_i$ can be seen as a multi-strategy $\multiStrat_i$ where, for all $hv \in \hist_i(\initVertex)$, $\multiStrat_i(hv)$ is the singleton $\{ \sigma_i(hv) \}$.
A \emph{multi-strategy profile} $\multiStrat = (\multiStrat_i)_{i \in \playerSet}$ is a tuple of multi-strategies, one per player.

Unlike strategies, when we fix a game $\game$ and a multi-strategy profile $\multiStrat$, there exist several plays beginning in $\initVertex$ that are consistent with all the multi-strategies~$\multiStrat_i$.
To describe them, we say that a strategy $\sigma_i$ is \emph{consistent} with a multi-strategy $\multiStrat_i$, written $\sigma_i \lesssim \multiStrat_i$ if for all $hv \in \hist_i(\initVertex)$, $\sigma_i(hv) \in \multiStrat_i(hv)$. We extend this notation to profiles of strategies, as expected. Then, we let $\multiOutcome{\multiStrat}{\initVertex}$ be the set of plays $\outcome{\sigma}{\initVertex}$ for all profiles $\sigma$ of strategies consistent with the multi-strategy $\multiStrat$. We call this set the \emph{outcomes of $\multiStrat$}. 
We also let $\multiHistory{\multiStrat}{\initVertex}$ be the set of histories consistent with the multi-strategy $\multiStrat$, \IE the finite prefixes of plays in $\multiOutcome{\multiStrat}{\initVertex}$.

\aline[inline]{Strongly et weakly étaient définis ici}

Our goal is to compute profiles of multi-strategies such that all profiles of consistent strategies are NEs or SPEs: such profiles of multi-strategies are called permissive NEs or permissive SPEs. By the existence of NEs and SPEs in reachability games, we straightforwardly obtain the existence of permissive NEs and permissive SPEs. We thus want to study \emph{most permissive} NEs or SPEs, \IE profiles of multi-strategies that are permissive NEs or SPEs, and such that no ``more permissive'' multi-strategies are still permissive NEs or~SPE{s}.

The natural first attempt would be to look for a notion of ``more permissive'' that is set-theoretic, with respect to a given solution concept. We would thus say that a profile of multi-strategies $\multiStrat$ is at least as permissive as a profile of multi-strategies $\multiStrat'$ if for all $i\in \playerSet$, for all histories $h\in \hist_i(\initVertex)$, $\multiStrat_i(h) \supseteq \multiStrat'_i(h)$. Then, $\multiStrat$ would be more permissive than $\multiStrat'$ if it is at least as permissive, while being different (for at most one history). Finally, $\multiStrat$ would be a \emph{most permissive} NE or SPE if it is a permissive NE or SPE, respectively, and no permissive NE or SPE, respectively, is more permissive than $\multiStrat$. 

\begin{figure}[ht]
\begin{center}
    \begin{tikzpicture}
    \node[draw, circle] (v0) at (0,0){$v_0$};
    \node[draw,circle, accepting] (v1) at (2,0){$v_1$};

    \draw[->] (v0) to [loop above] (v0);
    \draw[->] (v1) to [loop above] (v1);
    \draw[->] (v0) to (v1);

    \end{tikzpicture}
    \caption{In this game, player~$1$ owns all vertices and wants to reach $v_1$. For all $k \in \N$, we define the multi-strategy $\multiStrat^k_1$ such that for all $h \in \hist(\initVertex)$, $\multiStrat^k_1(h) = \{v_0,v_1 \}$ if $\last(h) = v_0$ and $|\{ n\in \N \mid h_n = v_0 \}| \leq k$, and $\multiStrat^k_1(h) = \{ v_1 \}$ otherwise. We have that for all $k \in \N$, for all $\sigma_1 \lesssim \multiStrat^k_1$, $\gain_1( \outcome{\sigma_1}{v_0})=1$ (and thus $\multiStrat^k_1$ is a permissive SPE), but for all $k \in \N$, $\multiOutcome{\multiStrat^{k}_1}{v_0} \subseteq \multiOutcome{\multiStrat^{k+1}_1}{v_0}$.}
    \label{fig:no-most-permissive-quali}
\end{center}
\end{figure}
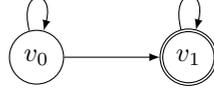

This natural definition is very problematic in the realm of reachability games (as already noticed in the context of winning strategies in parity games by \cite{BJW02}) where no most permissive NE or SPE could exist, as demonstrated by the game in Figure~\ref{fig:no-most-permissive-quali}. 

We thus propose another way to measure the permissiveness of a multi-strategy, inspired by the definition of penalty used in \cite{BDMR09} to describe permissive winning strategies in two-player games. To define the notion of penalty in our context, we equip the game with a function $\weight\colon E \to \N$ assigning a non-negative weight to each edge: if unspecified, we will consider that every edge has weight $1$. The player who owns the vertex at the source of an edge $e$ will pay the penalty $\weight(e)$ if he decides to not include the edge $e$ in his multi-strategy. All penalties are then counted additively. Formally, for a multi-strategy profile $\multiStrat$, we first define for each player $i \in \playerSet$ the \emph{penalty of player $i$ w.r.t.~$\multiStrat$} in a play $\rho = \rho_0\rho_1 \cdots$ by induction on the length of its prefixes:
\begin{itemize}
    \item $\penal{i}{\multiStrat}{\varepsilon} = 0 $ where $\varepsilon$ denotes the empty prefix;
    \item for $h = \rho_0\cdots\rho_k$,
    $\displaystyle \penal{i}{\multiStrat}{hv} =
    \begin{cases}\displaystyle \penal{i}{\multiStrat}{h} + \sum_{v' \in \successor(v) \setminus \multiStrat_i(hv)} \weight(v,v') & \text{ if } v \in V_i 
    \\ \penal{i}{\multiStrat}{h} & \text{ otherwise}
    \end{cases}$;
    \item $\penal{i}{\multiStrat}{\rho} = \lim_{k \rightarrow +\infty} \penal{i}{\multiStrat}{\rho_0\cdots\rho_k}$: this limit exists (it may be equal to $+\infty$) since $(\penal{i}{\multiStrat}{\rho_0\cdots\rho_k})_k$ is a non-decreasing sequence of natural numbers.
\end{itemize}

There are several ways to associate a penalty with a multi-strategy profile $\multiStrat$, depending on how we take into account the non-determinism offered in the multi-strategies. A first choice consists in considering a worst-case scenario in the outcomes (without considering the possible deviations). A second choice consists in  considering only the deviations of one player, \IE to consider that the retaliation of other players with respect to the deviation of a player will count in the final penalty. It is then possible to combine both types of penalties, though we will treat them separately in the rest of this article.

\begin{definition}[Penalties]
    Let $\multiStrat$ be a multi-strategy profile in $\initGame$. The \emph{main penalty} and \emph{retaliation penalty} of player $i$ with respect to $\multiStrat$ are defined respectively as \begin{align*}
      \mainPenalty_i(\multiStrat,\initVertex) &= \sup_{\rho\in \multiOutcome{\multiStrat}{\initVertex}} \penal i \multiStrat{\rho} 
    \\
    \retaliationPenalty_i(\multiStrat,\initVertex) &= \sup_{hv\in \hist_i(\initVertex)\setminus \multiHistory{\multiStrat}\initVertex}  \sup \{\penal i \multiStrat{\rho}\mid \rho \in \multiOutcome{\multiStrat_{\restriction hv}}{v} \}
    \end{align*}If there are no histories $hv$ in $\hist_i(\initVertex) \setminus \multiHistory{\multiStrat}{\initVertex}$, we let $\retaliationPenalty_i(\multiStrat,\initVertex) = 0$.
\end{definition}

%----Def strongly and weakly winning

The existence of a multi-strategy profile which satisfies some upper-bounds on penalties does not provide any certainty about the satisfaction of the reachability objectives of the players. For this reason, we also consider multi-strategy profiles that satisfy some properties on the set of players who satisfy their objective. Let $\winningPlayers$ be a subset of players  and $\multiStrat$ be a multi-strategy profile. Then, $\multiStrat$ is said \emph{weakly winning} if \textbf{there exists} a strategy profile $\sigma$ which is consistent with $\multiStrat$ and such its outcome is winning for all players in $\winningPlayers$. Similarly, $\multiStrat$ is said \emph{strongly winning} if \textbf{for each} strategy profile $\sigma$ which is consistent with $\multiStrat$, its outcome is winning for all players in $\winningPlayers$.

\begin{definition}[Weakly and strongly winning]
    Given a subset of player $\winningPlayers \subseteq \playerSet$ and a multi-strategy profile $\multiStrat$,
    \begin{itemize}
        \item  $\multiStrat$ is said \emph{weakly winning with respect to $\winningPlayers$} if there exists a strategy profile $\sigma$ such that $\sigma \lesssim \multiStrat$ and for all $i \in \winningPlayers$, $\gain_i(\outcome{\sigma}{\initVertex}) = 1$. 
\item $\multiStrat$ is said \emph{strongly winning with respect to $\winningPlayers$} if for all strategy profiles $\sigma$ such that $\sigma \lesssim \multiStrat$, we have that for all $i \in \winningPlayers$, $\gain_i(\outcome{\sigma}{\initVertex}) = 1$.
\end{itemize}
\end{definition}

%%%%%% Début du running example !

\begin{figure}[tbp]
\centering
\scalebox{0.9}{\begin{tikzpicture}
    \node[draw,circle] (v0) at (0,0){$v_0$};
    \node[draw,circle] (v2) at (1.5,-1){$v_2$};
    \node[draw,circle] (v1) at (1.5,1){$v_1$};
    \node[draw,circle,fill=gray!30] (v3) at (3,-1){$v_3$};
    \node[draw,accepting] (v4) at (3,1){$v_4$};

    \node[draw] (v5) at (-1.5,-1){$v_5$};
    \node[draw,circle,fill=gray!30,accepting] (v6) at (-1.5,1){$v_6$};
    \node[draw,circle] (v7) at (-3,0){$v_7$};
    \node[draw,circle,fill=gray!30] (v8) at (-4.5,1){$v_8$};
    \node[draw,circle,fill=gray!30] (v9) at (-4.5,-1){$v_9$};

    \draw[->] (v0) to (v1);
    \draw[->] (v0) to (v2);
    \draw[->] (v1) to (v4);
    \draw[->] (v1) to (v3);
    \draw[->] (v4) to (v3);
    \draw[->] (v2) to (v3);

    \draw[->] (v0) to (v5);
    \draw[->] (v5) to (v6);
    \draw[->] (v6) to (v0);
    \draw[->] (v5) to[loop right] node{$10$} (v5);
    \draw[->] (v3) to[loop right] (v3);
    \draw[->] (v5) to (v7);
    \draw[->] (v7) to (v8);

    \draw[->] (v7) to (v9);
    \draw[->] (v8) to[loop right] (v8);
    \draw[->] (v9) to[loop right] (v9);
\end{tikzpicture}
}
\caption{An example of a reachability game where player 1 (resp. player 2) owns circle (resp. rectangle) vertices. The initial vertex is $v_0$. Target vertices $F_{1} = \{ v_3,v_6,v_8,v_9 \}$ of player~$1$ and $F_{2} = \{v_4,v_6 \}$ of player~$2$  are drawn with gray vertices and double-bordered vertices respectively.} 
\label{fig:running-example1}
\end{figure}
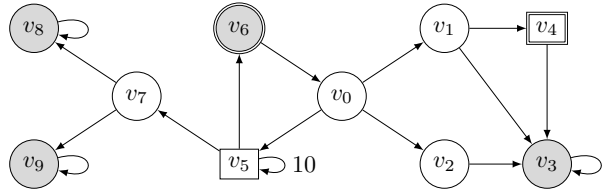

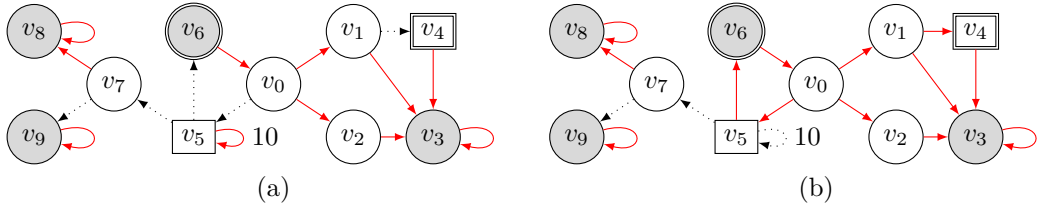
\begin{figure}[tbp]
\begin{subfigure}{0.49\textwidth}
\centering
    \scalebox{1}{\begin{tikzpicture}[scale=0.7]
    \node[draw,circle] (v0) at (0,0){$v_0$};
    \node[draw,circle] (v2) at (1.5,-1){$v_2$};
    \node[draw,circle] (v1) at (1.5,1){$v_1$};
    \node[draw,circle,fill=gray!30] (v3) at (3,-1){$v_3$};
    \node[draw,accepting] (v4) at (3,1){$v_4$};

    \node[draw] (v5) at (-1.5,-1){$v_5$};
    \node[draw,circle,fill=gray!30,accepting] (v6) at (-1.5,1){$v_6$};
    \node[draw,circle] (v7) at (-3,0){$v_7$};
    \node[draw,circle,fill=gray!30] (v8) at (-4.5,1){$v_8$};
    \node[draw,circle,fill=gray!30] (v9) at (-4.5,-1){$v_9$};

    \draw[->,red] (v0) to (v1);
    \draw[->,red] (v0) to (v2);
    \draw[->,dotted] (v1) to (v4);
    \draw[->,red] (v1) to (v3);
    \draw[->,red] (v4) to (v3);
    \draw[->,red] (v2) to (v3);

    \draw[->,dotted] (v0) to (v5);
    \draw[->,dotted] (v5) to (v6);
    \draw[->,red] (v6) to (v0);
    \draw[->, red] (v5) to[loop right] node{\color{black}$10$} (v5);
    \draw[->,red] (v3) to[loop right] (v3);
    \draw[->,dotted] (v5) to (v7);
    
     \draw[->,red] (v7) to (v8);

    \draw[->,dotted] (v7) to (v9);
    \draw[->,red] (v8) to[loop right] (v8);
    \draw[->,red] (v9) to[loop right] (v9);

        \node (a) at (0,-2){(a)};

\end{tikzpicture}
}
\end{subfigure}\hfill
\begin{subfigure}{0.49\textwidth}
\centering
    \scalebox{1}{\begin{tikzpicture}[scale=0.7]
    \node[draw,circle] (v0) at (0,0){$v_0$};
    \node[draw,circle] (v2) at (1.5,-1){$v_2$};
    \node[draw,circle] (v1) at (1.5,1){$v_1$};
    \node[draw,circle,fill=gray!30] (v3) at (3,-1){$v_3$};
    \node[draw,accepting] (v4) at (3,1){$v_4$};

    \node[draw] (v5) at (-1.5,-1){$v_5$};
    \node[draw,circle,fill=gray!30,accepting] (v6) at (-1.5,1){$v_6$};
    \node[draw,circle] (v7) at (-3,0){$v_7$};
    \node[draw,circle,fill=gray!30] (v8) at (-4.5,1){$v_8$};
    \node[draw,circle,fill=gray!30] (v9) at (-4.5,-1){$v_9$};

    \draw[->,red] (v0) to (v1);
    \draw[->,red] (v0) to (v2);
    \draw[->,red] (v1) to (v4);
    \draw[->,red] (v1) to (v3);
    \draw[->,red] (v4) to (v3);
    \draw[->,red] (v2) to (v3);

    \draw[->,red] (v0) to (v5);
    \draw[->,red] (v5) to (v6);
    \draw[->,red] (v6) to (v0);
    \draw[->,dotted] (v5) to[loop right] node{$10$} (v5);
    \draw[->,red] (v3) to[loop right] (v3);
    \draw[->,dotted] (v5) to (v7);
    \draw[->,red] (v7) to (v8);

    \draw[->,dotted] (v7) to (v9);
    \draw[->,red] (v8) to[loop right] (v8);
    \draw[->,red] (v9) to[loop right] (v9);

    \node (b) at (0,-2){(b)};
    
\end{tikzpicture}
}
\end{subfigure}
\caption{Examples of permissive equilibria: (a) a permissive NE and (b) a permissive SPE}
\label{fig:running-example2}
\end{figure}

\begin{example}
    An example of a reachability game with two players is depicted in Figure~\ref{fig:running-example1}. The edge labelled with $10$ corresponds to the penalty if player 2 decides not to allow this edge: all other penalties are set to $1$ by default. A multi-strategy is represented with red edges (black dotted edges are thus the ones that are not selected in the multi-strategy) in Figure~\ref{fig:running-example2}(a).\footnote{Notice that, in this example, the set of successors prescribed by  multi-strategies only depends on the current vertex and not on the past history.}
    All strategy profiles that are consistent with this multi-strategy depend on the choice of successor for $v_0$ among $\{v_1, v_2\}$.
    It is indeed a permissive NE since the consistent strategies are NEs: player 1 has no interest in deviating from either $v_1$ or $v_2$ in $v_0$, since all strategies lead to plays where he visits his target set, while going to $v_5$ make him lose.
    It has a main penalty of $2$ for player $1$ and $0$ for player $2$. Player $1$ can do slightly better by allowing the edge $(v_1, v_4)$ in the multi-strategy: this remains a permissive NE (now player $2$ wins in certain plays, but he is left with no real choices to make), and player 1 now gets a main penalty of $1$. This modified permissive NE is strongly winning w.r.t.~$\{1\}$, and weakly winning w.r.t.~$\{1, 2\}$. It is not a permissive SPE since player $2$ has a profitable deviation from $v_5$ by going to $v_6$ where he wins. A permissive SPE is depicted in Figure~\ref{fig:running-example2}(b), that is strongly winning w.r.t.~$\{1\}$, but only weakly winning w.r.t~$\{1, 2\}$. Player $2$ has a main penalty of $11$ (because he cuts edges $(v_5, v_5)$ and $(v_5, v_7)$), while player $1$ has a retaliation penalty of $1$ (because he cuts edge $(v_7, v_9)$). If we want a permissive SPE that is strongly winning w.r.t.~$\{1,2\}$, we need to increase the main penalty of player 1 to 2 by removing edges $(v_1,v_3)$ and $(v_0,v_2)$. However, we may decrease to 0 the retaliation penalty of player $1$ by adding the edge $(v_7,v_9)$ (since it is equally good to him anyway).
\end{example}

We now define the problems we study in the rest of the article, where we use the word ``equilibrium'' to either mean NE or SPE, depending on the solution concept we want to check. In all these problems, we give different penalty bounds for the main penalty and the retaliation penalty. Notice though that the bounds can be set to $+\infty$, relaxing the constraints in this case.

\begin{problem}[Constrained penalty problem]
Given a reachability game $\initGame$,  $m \in (\N \cup \{ \infty\})^n$ and $ r \in (\N \cup \{ \infty\})^n$, does there exist a permissive equilibrium $\multiStrat$ in $\initGame$ such that for all $i \in \playerSet$,
$\mainPenalty_i(\multiStrat,\initVertex) \leq m_i$ and $\retaliationPenalty_i(\multiStrat,\initVertex) \leq r_i$?
\end{problem}

\begin{problem}[Weakly winning with constrained penalty problem]
Given a reachability game $\initGame$,  $m \in (\N \cup \{ \infty\})^n$, $r \in (\N \cup \{ \infty\})^n$ and $\winningPlayers \subseteq \playerSet$,  does there exist a permissive equilibrium $\multiStrat$ in $\initGame$ such that \emph{(i)} for all $i \in \playerSet$,
$\mainPenalty_i(\multiStrat,\initVertex) \leq m_i$ and $\retaliationPenalty_i(\multiStrat,\initVertex) \leq r_i$ and \emph{(ii)} $\multiStrat$ is weakly winning w.r.t. $\winningPlayers$?   
\end{problem}

\begin{problem}[Strongly winning with constrained penalty problem]
Given a reachability game $\initGame$,  $m \in (\N \cup \{ \infty\})^n$, $r \in (\N \cup \{ \infty\})^n$ and $\winningPlayers \subseteq \playerSet$,  does there exist a permissive equilibrium $\multiStrat$ in $\initGame$ such that \emph{(i)} for all $i \in \playerSet$,
$\mainPenalty_i(\multiStrat,\initVertex) \leq m_i$ and $\retaliationPenalty_i(\multiStrat,\initVertex) \leq r_i$ and \emph{(ii)} $\multiStrat$ is strongly winning w.r.t. $\winningPlayers$? 
\end{problem}

We show in the rest of this article that all these problems, for NEs and SPEs, are decidable in PSPACE, if the upper-bound penalties are encoded in unary. To do so, we characterize the permissive equilibria in the various problems in Section~\ref{section:charac}. In Section~\ref{section:comput}, we then show that tree-like witnesses can be found if the according permissive equilibria exist. These witnesses have a height bounded by a polynomial depending on the size of the game and the largest upper-bound on penalties. We use these witnesses to obtain the PSPACE decision procedures.

%---
% Charac
%---

\section{Characterizations of permissive equilibria}
\label{section:charac}

We now characterize permissive equilibria of the reachability game $\initGame$. This is a first step towards their computation in the next section.  
We provide a characterization for permissive NEs in Section~\ref{section:NEcharac} and one for permissive SPEs in Section~\ref{section:SPEcharac}.
These characterizations are inspired by existing ones for classical  NEs (resp.~SPEs)~\cite{BrihayeBGT19,weakSPEs}. The latter rely on properties that a play (resp.~a set of plays) must satisfy in order to be the outcome of an NE (resp.~the set of subgame outcomes of an SPE). However, the outcomes of permissive equilibria are a set of plays and not a simple play. For that reason, the characterizations of permissive equilibria employ trees that we first formally define.

\subparagraph*{Trees} We call \emph{tree over $\game$ rooted at $v$} (for some $v \in V$) any subset $\tree$ of non-empty histories of $\game$ that contains $v$ and such that if $hu \in \tree$ then $h\in \tree$.
All $h \in \tree$ are called \emph{nodes} of the tree, the particular node $v$ is called the \emph{root} of the tree, and for all $hu \in \tree$, $h$ is called the \emph{parent} of $hu$, and $hu$ a \emph{child} of $h$.

As for histories in an arena, for all $hu \in \tree$, we let $\last(hu) = u$.
The \emph{depth} of a node $h \in \tree$, written $\depth(h)$, is equal to $|h|$ and its \emph{height}, denoted by $\height(h)$, is given by $ \sup \{ |\last(h)h'| \mid h' \in \hist \text{ and } h h' \in \tree \}$.
The height of the tree corresponds to the height of its root.
A node $h \in \tree$ is called a \emph{leaf} if $\height(h) =0$.

We denote by $\tree_{\restriction hu}$, the subtree of $\tree$ rooted at $u$ for some $hu \in \tree$, that is the set of non-empty histories $h'\in \hist(u)$ such that $hh' \in \tree$.

A (finite or infinite) \emph{branch} of the tree is a maximal (finite or infinite) sequence of nodes $h_0h_1\cdots$ such that for all $k \in \N$, $h_{k}$ is the parent of $h_{k+1}$.
Finally, we denote by $\infTree$ the set of plays in $\game$ represented by infinite branches in $\tree$, \IE 
\[\infTree = \{ \rho_0\rho_1 \cdots \in \plays \mid \text{there exists a branch } h_0h_1 \cdots \in \tree \text{ st. } \forall k \in \N, \;  \rho_k = \last(h_k) \}\]

\smallskip
In what follows, we consider outcomes of multi-strategies as trees. Indeed, given a multi-strategy $\multiStrat$, $\multiHistory{\multiStrat}{\initVertex}$ can be seen as a tree $\tree$ over $\game$ rooted at $v_0$ and $\multiOutcome{\multiStrat}{\initVertex}$ corresponds to $\infTree$. 
In particular, penalties can also be defined on trees, mimicking the definition for profiles of multi-strategies. The penalty of a tree $\tree$ for a player $i$, denoted by $\Penalty_i(\tree)$, is the maximal penalty of a branch of $\tree$, the penalty of a branch being equal to the penalty of the associated play $\rho$ w.r.t.~any profile of multi-strategies that is consistent with the choices appearing in $\tree$ along the play $\rho$. Formally,  let $\tree$ be a tree and $i \in \playerSet$ be a player.
For each $hv = v_1\cdots v_k v\in \tree$, we define $\blocked(h) = \{ u \in \successor(v_k) \mid hu \not \in \tree \} $ as the set of blocked successors of $h$ in $\tree$ and  \[\penalAux_i(hv) = \begin{cases} 0  & \text{ if } h= \varepsilon \\
    \penalAux_i(h) + \sum_{u \in \blocked(h)} w(v_k,u) & \text{ if } v_k \in V_i \\
     \penalAux_i(h) & \text{ otherwise.}\end{cases}\]
Moreover, for all plays $\rho=\rho_0\rho_1\cdots \in \infTree$, we let $\penalAux_i(\rho) = \lim_{k \rightarrow +\infty} \penalAux_i(\rho_0 \cdots \rho_k)$.
Thus, the penalty of a tree $\tree$ for a player $i \in \playerSet$ is naturally defined as: 
\[\penalAux_i(\tree) =  \sup \{ \penalAux_i(\rho) \mid \rho \in \infTree \}.\]

%---
% NE charac
%---

\subsection{Characterization of permissive Nash equilibria}
\label{section:NEcharac}

In order to characterize permissive Nash equilibria, we start by defining \emph{good} trees, by checking two conditions. The first one, called \emph{resistance to internal deviations}, means that at any node~$h$ of the tree such that $\last(h)$ belongs to player $i$, if $h$ has at least two children, the plays starting with $h$ are either all losing, or all winning, for player~$i$. The second one, called \emph{resistance to external deviations}, means that at any node $hu$ of the tree with $u$ belonging to player $i$, if player $i$ has the possibility to play to a successor $u'$ not in the tree from which he has a winning strategy, then all plays in the subtree from $hu$ must be winning for player~$i$.

\begin{definition}
    \label{def:goodtree}

    Let $\tree$ be a tree over $\initGame$. 
    \begin{enumerate}

     \item Given a subset of players $\deviators \subseteq \playerSet$, the tree $\tree$ is \emph{$\deviators$-resistant to internal deviations} if for all $i \in \deviators$ and for all $hv \in \tree$ such that $v \in V_i$ and $|\{hvv'\in \tree\mid v' \in V\}| \geq 2$, we have that  for all $\rho, \rho'\in \tree^\infty_{\restriction hv}$, $\gain_i(h\rho) = \gain_i(h\rho')$. If $\deviators = \playerSet$, we simply say that $\tree$ is \emph{resistant to internal deviations}. \label{def:internalDev}
    \item The tree $\tree$ is \emph{resistant to external deviations} if for all $hu \in \tree$ with $u \in V_i$ and $i \not \in \visit(hu)$, if there exists $u'\in \successor(u)$ such that $huu'\notin \tree$ and player $i$ has a winning strategy from $u'$ (against the coalition of the other players), then for all plays $\rho \in \tree^\infty_{\restriction hu}$, $\gain_i(\rho)=1$.
    \item The tree $\tree$ is \emph{good} if it is resistant to internal and external deviations. 
    \end{enumerate}
\end{definition}

The resistance to internal and external deviations leads to the characterization of outcomes of permissive NEs (Theorem~\ref{thm:NECharac}): given a good tree $\tree$, there exists a permissive NE such that its outcomes are the plays corresponding to the infinite branches of $\tree$ iff $\tree$ is good.

\begin{restatable}{theorem}{thmNECharac}
\label{thm:NECharac}
Let $\tree$ be a tree over $\initGame$ rooted at $v_0$. The following assertions are equivalent:
\begin{enumerate}
    \item There exists a permissive NE $\multiStrat$ in $\initGame$ such that $\multiHistory{\multiStrat}{\initVertex} = \tree$;
    \item The tree $\tree$ is good.
\end{enumerate}
\end{restatable}

\begin{remark}\label{rem:penalties}
    For all multi-strategies $\multiStrat$, and all players $i\in \playerSet$, the penalty $\mainPenalty_i(\multiStrat, \initVertex)$ is equal to the penalty of player $i$ in the good tree $\multiHistory{\multiStrat}{\initVertex}$, \IE $\Penalty_i(\multiHistory{\multiStrat}{\initVertex})$. The construction of Theorem~\ref{thm:NECharac} thus also preserves the main penalties. 
\end{remark}

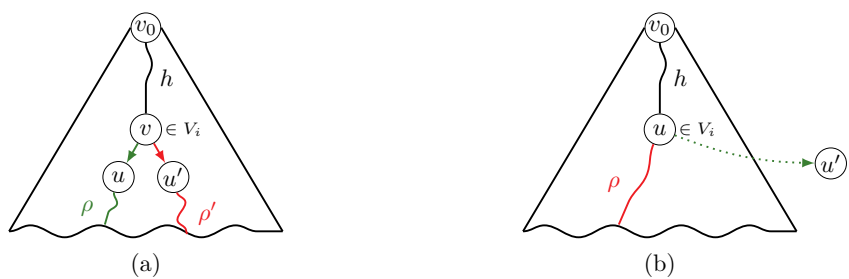
\begin{figure}
    \centering
    \begin{subfigure}[t]{0.49\textwidth}
    \centering\scalebox{.9}{
        \begin{tikzpicture}[>=latex]
    \node[draw,circle,inner sep=0.5pt] (v) at (0,0){$v_0$};
    \draw[thick,-] (v.west) to (-2,-3);
    \draw[thick,-] (v.east) to (2,-3);
    \draw[thick,decorate,decoration=snake, segment length=1cm] (-2,-3) to (2,-3);

    \node[draw,circle,inner sep=0.5pt, minimum size=13pt] (h) at (0,-1.5){$v$};
    \node (Vi) at (0.55,-1.5){\scriptsize$\in V_i$};
        \draw[thick,decorate,decoration=snake, segment length=1cm] (v) to node[right=2pt]{$h$} (h);

        \node[draw,circle, inner sep=0.5pt, minimum size=13pt] (u) at (-0.4,-2.2){$u$};

        \node[draw,circle, inner sep=0.5pt,  minimum size=13pt] (u') at (0.4,-2.2){$u'$};

        \draw[->,thick, color=OliveGreen] (h) -- (u);

        \draw[->,thick, color=Red] (h) -- (u');

        \draw[thick,decorate,decoration=snake, segment length=0.5cm, color=OliveGreen] (u)  to node[left=3pt]{\color{OliveGreen}$\rho$} (-0.6,-2.9);
        \draw[thick,decorate,decoration=snake, segment length=0.5cm, color=Red] (u') to node[right=3pt]{\color{Red}$\rho'$} (0.6,-3.04);

        \node() at (0,-3.5) {(a)};

\end{tikzpicture}}
    \end{subfigure}\hfill
    \begin{subfigure}[t]{0.49\textwidth}
    \centering\scalebox{.9}{
    \begin{tikzpicture}[>=latex]
    \node[draw,circle,inner sep=0.5pt] (v) at (0,0){$v_0$};
    \draw[thick,-] (v.west) to (-2,-3);
    \draw[thick,-] (v.east) to (2,-3);
    \draw[thick,decorate,decoration=snake, segment length=1cm] (-2,-3) to (2,-3);

    \node[draw,circle,inner sep=0.5pt, minimum size=13pt] (h) at (0,-1.5){$u$};
    \node (Vi) at (0.55,-1.5){\scriptsize$\in V_i$};
        \draw[thick,decorate,decoration=snake, segment length=1cm] (v) to node[right=2pt]{$h$} (h);

        \node[draw,circle, inner sep=0.5pt,  minimum size=13pt] (u') at (2.5,-2){$u'$};

        \draw[->,thick, color=OliveGreen,dotted] (h) edge[bend right=10] (u');

        \draw[thick,decorate,decoration=snake, segment length=1.5cm, color=Red] (h)  to node[left=3pt]{\color{Red}$\rho$} (-0.6,-2.9);
    
\node() at (0,-3.5) {(b)};
        
\end{tikzpicture}   }
    \end{subfigure}
    \caption{Examples of trees that do not respect: (a) the resistance to internal deviations since $\gain_i(\rho')=0$ but $\gain_i(\rho) = 1$; (b) the resistance to external deviations since $\gain_i(\rho) = 0$ but Player~$i$ can win from $u'$.}
    \label{fig:good-tree}
\end{figure}

\begin{proof}[Proof sketch]

For $(1 \Rightarrow 2)$ let us assume that $\multiStrat$ is a permissive NE and that $\multiHistory{\multiStrat}{\initVertex} = \tree$.
We have to prove that $\tree$ is good.
If $\tree$ is not resistant to internal deviations that means that from some vertex $v$ there exists two plays $\rho$, crossing $u$, and $\rho'$, crossing $u' \neq u$, such that: $\rho$ is winning for player~$i$ and $\rho'$ is losing for player~$i$, see Figure~\ref{fig:good-tree}(a). In particular, we can build a strategy profile $\sigma$ consistent with $\multiStrat$ such that $\outcome{\sigma_{\restriction h}}{v} = \rho'$ and $\outcome{\sigma_{\restriction hv}}{u} = \rho_{\geq 1}$. Meaning that player~$i$ should deviate by choosing $u$ instead of $u'$ from $v$, meaning that  $\sigma$ is not an NE and $\multiStrat$ is not a permissive NE.
If $\tree$ is not resistant to external deviations, that means that from some vertex $u$ of player~$i$ there exists a play $\rho$ such that $\gain_i(\rho) = 0$ and $u'$ a successor of $u$ outside $\tree$ from which player~$i$ can win, see Figure~\ref{fig:good-tree}(b). Thus we can build a strategy profile $\sigma$ consistent with $\multiStrat$ such that $\outcome{\sigma}{\initVertex} = h\rho$. In this way, player~$i$ should choose to go in $u'$ and then follow a winning strategy meaning that $\sigma$ is not an NE and $\multiStrat$ not a permissive NE. 

For $(2\Rightarrow 1)$, let us assume that $\tree$ is a good tree. We build a permissive NE $\multiStrat$ such that its outcomes are the plays corresponding to the infinite branches of $\tree$. Additionally, if a player~$i$ deviates from $\tree$, the coalition of the other players plays its retaliation\footnote{This retaliation strategy corresponds to the winning strategy of player~$2$ in a two-player zero-sum reachability game in which player~$1$ is player~$i$ and wants to reach $\targetSet_i$ and player~$2$ is the coalition of the other players and wants to avoid visiting $\targetSet_i$ \cite[Chapter 2]{lncs2500}.} strategy to prevent player~$i$ from deviating. \end{proof}

%---
% SPE charac
%---

\subsection{Characterization of permissive subgame perfect equilibria}
\label{section:SPEcharac}

Permissive subgame perfect equilibria are intrinsically more complex than permissive Nash equilibria. Thus their characterization cannot only rely on the outcomes from the initial vertex, it should also take into account the outcomes in all subgames.  This is the reason why, in order to deal with a compact representation of outcomes of a permissive SPE and its subgames, we introduce the notion of \emph{forest}. Then, we generalize the definition of good trees to define \emph{good forests} needed to characterize SPEs instead of NEs.

\subparagraph*{Forests and penalties of forests}  Trees of the forest are indexed by tuples $(i,v,\I) \in \playerSet \times V\times 2^{\playerSet}$. More precisely, we let 
\[\mathcal{I} = \{ (0,\initVertex,\initVisit) \} \cup \{ (i,v,\I) \in \playerSet \times V \times 2^{\playerSet} \mid \exists hv' \in \hist_i(\initVertex) \text{ st. } v \in \successor(v') \, \wedge \, \I = \visit(hv'v) \}\]
where $\initVisit = \{i\in \playerSet\mid \initVertex\in F_i\}$. Apart from the special tuple $(0,\initVertex,\initVisit)$, a tuple $(i,v,\I)$ represents the fact that $v$ is a vertex played by player~$i$ and reachable from $\initVertex$, and that all players in $\I$ have already seen their target when $v$ is reached. 
A forest in $\initGame$ is thus a set of trees $\forest = \{ \tree_{i,v,I} \mid (i,v,\I) \in \mathcal{I}\}$ such that $\tree_{i,v,\I}$ is a tree without leaves over $\game$ rooted at $v$.
The intuition behind this object is that the tree $\tree_{0,\initVertex,\initVisit}$ represents the outcomes of a multi-strategy $\multiStrat$ and the other trees $\tree_{i,v,\I}$ represent the outcomes of 
$\multiStrat_{\restriction hv'}$ in the subgames $(\game_{\restriction hv'},v)$ 
for all $hv' \in \hist_i(\initVertex)$ such that $\visit(hv'v) = \I$.

Moreover the main (resp.~retaliation) penalty of a forest $\forest$ for a player $i \in \playerSet$ are  respectively given by 
$$ \mainPenalty_i(\forest) = \penalAux_i(\tree_{0,\initVertex,\initVisit}) \quad\text{ and } \quad
 \retaliationPenalty_i(\forest) \displaystyle = \sup_{\substack{ \tree_{i,v,I} \in \forest \\ (i,v,\I) \in \Out  }} \!\!\!\! \penalAux_i(\tree_{i,v,I})$$
 where $\Out = \{(i,v,I) \in \mathcal{I} \setminus \{ (0,\initVertex,\initVisit) \} \mid \exists hv \in \hist(\initVertex) \text{ st. } hv \not \in \tree_{(0,\initVertex,\initVisit)} \, \wedge \, \last(h) \in V_i \wedge \I = \visit(hv) \}$ described the indices of trees in the forest that are deviations from the main tree $\tree_{0,\initVertex,\initVisit}$. If $\Out$ is empty, we let $\retaliationPenalty_i(\forest) = 0$.

\subparagraph{Characterization} 
Following the same philosophy as for permissive NEs, a forest is \emph{good} if  each tree $\tree_{i,v,\I}$ of the forest satisfies two properties. 
The first one is that $\tree_{i,v,\I}$ has to be $(\playerSet\setminus \I)$-resistant to internal deviations, exactly as for permissive NEs except that we take into account players who have already visited their target set, \IE players in $\I$.
The second one, called \emph{resistance to constrained external deviations}, means that at any node $hu$ of the tree such that $u$ belongs to player~$j$, if player~$j$ has the possibility to jump to another tree $\tree_{j,u',\I'}$ by playing to a successor $u'$ not in the tree and if there exists a play in this latter tree which is winning for player~$j$, then all plays after $hu$ in $\tree_{i,v,\I}$ have to be winning for player~$j$.

\begin{definition}[Good forest]
    \label{def:goodWitness}

    Let $\forest$ be a forest in $\initGame$.
    \begin{enumerate}
       \item A tree $\tree_{i,v,\I} \in \forest$ is \emph{resistant to constrained external deviations} if it satisfies the following property: for all $hu \in  \tree_{i,v,\I}$ and $j \in \playerSet$ such that we have that \emph{(i)} $u \in V_j$ and $j \not \in I \cup \visit(hu)$ and \emph{(ii)} there exists $u' \in \successor(u)$ such that $huu' \not \in \tree_{i,v,\I}$, if there exists $\rho' \in \tree^\infty_{j,u',\I'}$, where $\I' = \I \cup \visit(huu')$, such that $\gain_j(\rho') = 1$, then for all $\rho \in \tree^\infty_{i,v,\I \restriction hu}$,  $\gain_j(\rho) = 1$.

        \item The forest $\forest$ is \emph{good} if each tree $\tree_{i,v,\I} \in \forest$ is $(\playerSet \setminus \I)$-resistant to internal deviations (see~\eqref{def:internalDev} in Definition~\ref{def:goodtree})  and  resistant to constrained external deviations. 
    \end{enumerate}
    \end{definition}

Thanks to good forests, we are able to characterize the outcomes of permissive SPEs: given a good tree $\tree^*$, there exists a permissive SPE such that its outcomes correspond to $\tree^*$ iff there exists a good forest whose ``main'' tree is $\tree^*$, \IE $\tree_{0,\initVertex,\initVisit} = \tree^*$. With some other constraints, this also preserves strongly (resp. weakly) winning and penalty properties.

\begin{restatable}{theorem}{thmSPECharac}
\label{thm:SPECharac}
Let $m \in (\N \cup \{ \infty \})^n$ and $r \in (\N \cup \{ \infty \})^n$ be upper thresholds. Let $\tree^*$ be a tree rooted at $\initVertex$ and $\winningPlayers \subseteq \playerSet$ be a set of players. The following assertions are equivalent:
\begin{enumerate}
    \item There exists a permissive SPE $\multiStrat$ in $\initGame$ such that:
    \begin{enumerate}
        \item $\multiHistory{\multiStrat}{\initVertex} = \tree^*$;
        \item $\multiStrat$ is strongly winning w.r.t. $\winningPlayers$; \label{assert:strongWinStrat}
        \item for all $i \in \playerSet$, $\mainPenalty_i(\multiStrat,\initVertex) \leq m_i$ and $\retaliationPenalty_i(\multiStrat,\initVertex) \leq r_i$.
    \end{enumerate}
    \item There exists a good forest $\forest$ in $\initGame$ such that:
    \begin{enumerate}
        \item $\tree_{0,\initVertex, \initVisit} = \tree^*$;
        \item for all $\rho \in \tree^\infty_{0,\initVertex,\initVisit}$, for all $i \in \winningPlayers$, $\gain_i(\rho) = 1$; \label{assert:strongWinWitness}
        \item for all $i \in \playerSet$, $\mainPenalty_i(\forest) \leq m_i$ and $\retaliationPenalty_i(\forest) \leq r_i$.
    \end{enumerate}
\end{enumerate}
These assertions are still equivalent by replacing \ref{assert:strongWinStrat} by ``$\multiStrat$ is weakly winning w.r.t.~$\winningPlayers$'' and \ref{assert:strongWinWitness} by ``there exists $\rho \in \tree^\infty_{0,\initVertex,\initVisit}$ such that for all $i \in \winningPlayers$, $\gain_i(\rho) = 1$''.
\end{restatable}

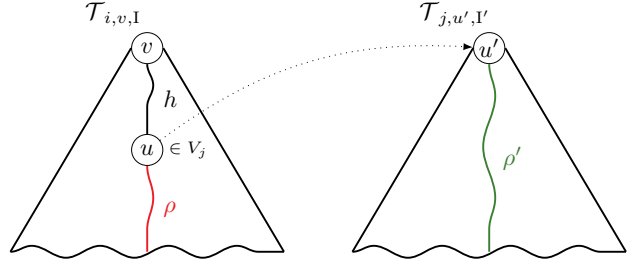
\begin{figure}[tbp]
\centering\scalebox{.9}{
\begin{tikzpicture}

   \node[draw,circle,inner sep=0.5pt,minimum size=13pt] (v) at (0,0){$v$};
   \node (tree1) at (-0.5,0.5){$\tree_{i,v,\I}$};
    \draw[thick,-] (v.west) to (-2,-3);
    \draw[thick,-] (v.east) to (2,-3);
    \draw[thick,decorate,decoration=snake, segment length=1cm] (-2,-3) to (2,-3);

    \node[draw,circle,inner sep=0.5pt,minimum size=13pt] (h) at (0,-1.5){$u$};
    \node (Vi) at (0.6,-1.5){\scriptsize$\in V_j$};
        \draw[thick,decorate,decoration=snake, segment length=1cm] (v) to node[right=3pt]{$h$} (h);

        \draw[thick,decorate,decoration=snake, segment length=1.5cm, color=Red] (h) to node[right=3pt]{\color{Red} $\rho$} (0,-3);

%Second tree
    \begin{scope}[yshift=1.5cm]
    \node[draw,circle,inner sep=0.5pt,minimum size=13pt] (u) at (5,-1.5){$u'$};
    \node (tree2) at (4.5,-1){$\tree_{j,u',\I'}$};
    \draw[thick,-] (u.west) to (3,-4.5);
    \draw[thick,-] (u.east) to (7,-4.5);
    \draw[thick,decorate,decoration=snake, segment length=1cm] (3,-4.5) to (7,-4.5);

        \draw[->,>=latex,dotted] (h) edge[bend left=20] (u);
        \draw[thick,decorate,decoration=snake, segment length=1.5cm, color=OliveGreen] (u) to node[right=2pt]{\color{OliveGreen}$\rho'$} (5,-4.5);
    \end{scope}
        
\end{tikzpicture}}
\caption{Example of forest that does not respect the resistance to constrained external deviations since $\gain_i(\rho)=0$ but $\gain_i(\rho')=1$.}
\label{fig:good-forest}
\end{figure}

\begin{proof}[Proof sketch]For $(1 \Rightarrow 2)$,
let us assume that $\multiStrat$ is a permissive SPE.
We build a good forest $\forest$ such that $\tree_{0,\initVertex,\initVisit}$ is the outcomes of $\multiStrat$, \IE  $\tree_{0,\initVertex,\initVisit}= \multiHistory{\multiStrat}{\initVertex}$, and a tree $\tree_{i,v,\I}$ is a representative of the outcomes of $\multiStrat_{\restriction hv'}$ in some subgame $(\game_{\restriction hv'}, v)$ such that $v' \in V_i$ and $\visit(hv'v) = \I$. In order to obtain a good forest and since several $hv'v$ could satisfy those properties, each representative $\tree_{i,v,\I}$ has to be chosen in a proper way: it has to minimize the maximal gain of player~$i$ for plays in  $\tree_{i,v,\I}$. More formally, 
for each $(i,v, \I) \in \mathcal{I}$, we let
        $\mathcal{O}(i,v, \I) =  \{ \multiHistory{\multiStrat_{\restriction hv'}}{v} \mid hv'v \in \hist(\initVertex)  \, \wedge \,v' \in V_i \, \wedge \, \I = \visit(hv'v) \}$ and we choose  $\tree_{i,v,\I} \in \mathcal{O}(i,v,\I)$ such that
        $
            \max\{ \gain_i(\rho) \mid \rho \in \tree^\infty_{i,v, \I} \} = \min_{\tree \in \mathcal{O}(i,v,\I)} \max \{ \gain_i(\rho) \mid \rho \in \infTree\}$.

Thanks to this latter property, $\forest$ is good. Indeed, let $\tree_{i,v,\I}$ be a tree of $\forest$.
Exactly as for permissive NEs, if $\tree_{i,v,\I}$ is not $(\playerSet\setminus \I)$-resistant to internal deviations, we can build a strategy profile $\sigma$ consistent with $\multiStrat$ such that the restriction of $\sigma$ is not an NE in a subgame corresponding to $\tree_{i,v,\I}$. If $\tree_{i,v,\I}$ is not resistant to constrained external deviations that means that from some node $u$, owned by player~$j$, there exists a play $\rho$ losing for player~$j$ and player~$j$ could choose to play outside $\tree_{i,v,\I}$ by jumping to a tree $\tree_{j,u',\I'}$ in which there exists a play $\rho'$ winning for him, see Figure~\ref{fig:good-forest}. 
Let $g$ be the history such that $\tree_{i,v,\I}$ represents the outcomes of $\multiStrat_{\restriction g}$ in $(\game_{\restriction g},v)$. Notice that $\outcome{\multiStrat_{\restriction gh}}{u'}$ may be different from $\tree_{j,u',I'}$. However thanks to the way in which this representative is chosen, we have that there exists a play $\rho''$ in $\outcome{\multiStrat_{\restriction gh}}{u'}$ with $\gain_j(\rho'') = 1$. Thus, we can build a strategy $\sigma$ consistent with $\multiStrat$ such that $\outcome{\sigma_{\restriction gh}}{u} = \rho$ and $\outcome{\sigma_{\restriction ghu}}{u'} = \rho''$. This means that player~$j$ could deviate by choosing $u'$ instead of $u$ from $v$ in the subgame $(\game_{\restriction g},v)$, thus $\sigma$ would not be an SPE and $\multiStrat$ not a permissive SPE.

For $(2 \Rightarrow 1)$, from a good forest $\forest$ a multi-strategy is build such that its subgame outcomes are the trees of $\forest$. This forms a permissive SPE because $\forest$ is good.
\end{proof}

For now, good trees and trees in good forests are infinite, but Section~\ref{section:comput} will show that we can represent some trees using a finite representation (intuitively, by supposing that every branch ends with a lasso in the game). It is this finite representation of good trees and good forests that will be used to decide the constrained penalty problems for  permissive NEs and permissive SPEs, thanks to the characterizations of Theorems~\ref{thm:NECharac} and~\ref{thm:SPECharac}.

%---
% Computation
%---

\section{Computation of permissive equilibria}
\label{section:comput}

Theorems~\ref{thm:NECharac} and~\ref{thm:SPECharac} characterize permissive NEs and SPEs with respect to infinite tree-shaped objects. In this section, we use these characterizations in order to decide the various penalty problems defined in Section~\ref{sec:permissiveness}: we check the existence of the good infinite tree-shaped objects by checking the existence of finite symbolic representations of such objects. We start by describing for a single tree this symbolic representation, and show that there exists a \emph{polynomial-size} such representation (when the penalty upper-bounds are encoded in unary).

\subsection{Symbolic trees and forests}

\newcommand\stree{\mathcal U}

\begin{definition}
    A \emph{symbolic tree} is a pair $\stree = (\tree, f)$ with $\tree$ a finite tree (\IE a finite subset of non-empty histories of $\game$), and $f$ a function mapping each leaf $h$ of $\stree$ to a non-empty set of successor nodes $h'$ that are ancestors of $h$ in $\stree$ such that $(\last(h), \last(h')) \in E$.
\end{definition}

A symbolic tree can be \emph{unfolded} into an infinite tree by repeatedly expanding the leaves of $\stree$ using as successors the choice prescribed by $f$. We denote by $\widetilde \stree$ the infinite tree obtained by unfolding the symbolic tree $\stree$. Similarly, the notions of symbolic forest $\forest$, where every tree in it is a symbolic tree, and unfolding of symbolic forest $\widetilde\forest$ can be defined.

In order to treat simultaneously NEs and SPEs, we introduce a new definition generalizing the resistance to external deviations and constrained external deviations.
For a vector $\gamma \in \{0, 1\}^{\playerSet\times V\times 2^{\playerSet}}$ of gains, and a subset $D\subseteq \playerSet$ of players (that represent players that did not already win at the beginning of the tree), we say that a tree $\tree$ is $(\gamma, D)$-resistant if for all $hu\in \tree$ with $u\in V_i$ and $u'\in \successor(u)$ with $huu'\notin \tree$, if $\gamma_{i, u', (\playerSet\setminus D)\cup \visit(huu')} = 1$, if $i \not \in (\playerSet\setminus D)\cup \visit(hu) $,  then for all plays $\rho \in \tree^\infty_{\restriction hu}$, $\gain_i(\rho)=1$. 

\begin{remark}\label{rem:gamma}
    The notion of $(\gamma,D)$-resistance is close to the resistance to external deviations and constrained external deviations, so that we directly obtain from Theorems~\ref{thm:NECharac} and \ref{thm:SPECharac}:
    \begin{itemize}
        
        \item Let $\gamma^{\game}$ be defined as follows: for all $(i, u, \I)$, we let $\gamma^{\game}_{i, u, \I}$ equals $1$ iff player $i$ belongs to $\I$ or can win from $u$ against the coalition of the other players in $\game$. Let $\tree$ be a tree. Then, $\tree$ is a good tree iff $\tree$ is resistant to internal deviations and $(\gamma^{\game}, \playerSet)$-resistant.
        
        \item Let $\forest$ be a forest and let $\gamma^{\forest}$ defined as follows: for all $(j, u, \J)$, we let $\gamma^{\forest}_{j, u, \J}$ equals $1$ iff player $j$ belongs to $\J$ or the tree $\tree_{j, u, \J}$ contains at least one branch with a vertex of $F_j$. Then, $\forest$ is a good forest iff each  each tree $\tree_{i, v, \I}$ of $\forest$ is $(\playerSet \setminus \I)$-resistant to internal deviations and $(\gamma^{\forest}, \playerSet \setminus \I)$-resistant.
    \end{itemize}
\end{remark}

The challenge to make this remark a decision procedure is to make the tree and forest finitely representable. We treat each tree independently of each other, thus explaining how to symbolically represent one single tree in the following proposition:
\newcommand\none{\bot}

\begin{restatable}{proposition}{bound}
%\begin{proposition}
\label{prop:bound}
    Let $\tree$ be a tree that is $D$-resistant to internal deviations, with $D\subseteq \playerSet$. We let $\gamma \in \{0, 1\}^{\playerSet\times V\times 2^{\playerSet}}$ be a vector of gains such that $\tree$ is $(\gamma,D)$-resistant, and $(P_i)_{i\in \playerSet'}$ be finite constraints on penalties for a subset $\playerSet'\subseteq \playerSet$ of players. There exists a symbolic tree~$\stree$, that is a subtree of $\tree$, of height polynomial in the number of players and vertices of $\game$, and in the largest bound on penalty $P_i$, such that the infinite tree $\widetilde \stree$ satisfies the following properties:
    \begin{enumerate}
        \item $\widetilde \stree$ is $D$-resistant to internal deviations;
        \item in $\widetilde \stree$, every player $i\in \playerSet'$ has a penalty at most $P_i$;
        \item $\widetilde \stree$ is $(\gamma,D)$-resistant.
    \end{enumerate}
    Moreover, for a subset $\winningPlayers$ of players, 
    if we start with $\tree$ that is strongly (respectively, weakly) winning w.r.t.~$\winningPlayers$, then we can make the above construction so that moreover $\widetilde \stree$ is strongly (respectively, weakly) winning w.r.t.~$\winningPlayers$. 
%\end{proposition}
\end{restatable}

    \begin{figure}[tbp]
        \centering
        \includegraphics[height=3cm]{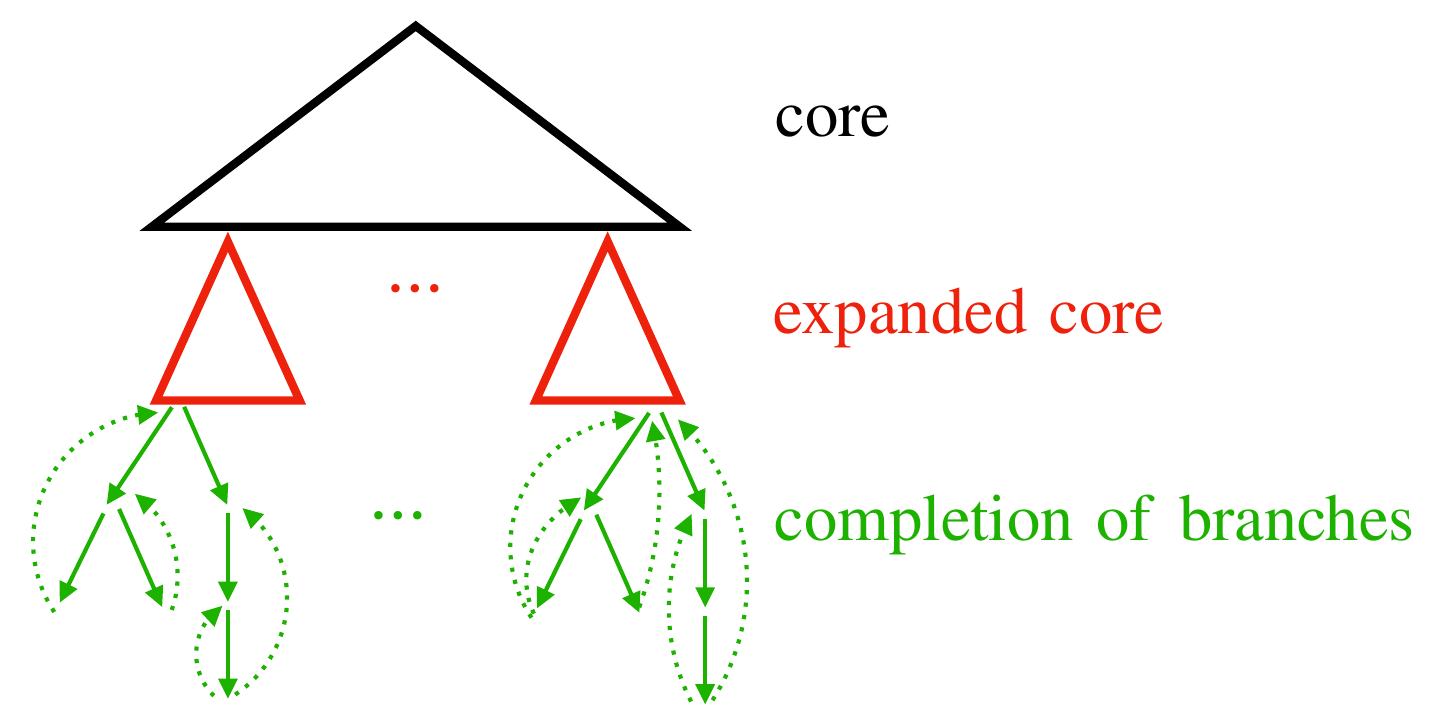}
        \caption{Construction of the symbolic tree}
        \label{fig:symbolic-tree-body}
    \end{figure}
    
The proof of this result goes by several steps, that we briefly sketch here only in the case where $\tree$ is strongly winning w.r.t.~$\winningPlayers$. Figure~\ref{fig:symbolic-tree-body} depicts the notions used in the construction of the symbolic tree. First, we consider the smallest subtree of $\tree$ where leaves are such that all players of $\winningPlayers$ have visited their target set: this subtree is finite by König's lemma, since all branches of $\tree$ have such a node where all players of $\winningPlayers$ have won, and the tree is finitely branching. This subtree is called the \emph{core}. We then continue considering the parts of $\tree$ outside the core, in order to complete the branches so that: the $D$-resistance to internal deviations is fulfilled (if a player has won in a certain branch of a subtree, he must win in all of them), the $(\gamma, D)$-resistance is fulfilled (if $\gamma$ gives a constraint in the current node for a player $i$, all the branches of this subtree should visit a target vertex of $i$). This extension of the core is cut into two parts: the \emph{expanded core} that ends in places where all the new players that must visit their target because of $D$-resistance to internal deviations and $(\gamma, D)$-resistance have indeed won; the \emph{completion of branches} in order to then find leaves of the symbolic tree where all successors can be replaced (with function $f$) by similar nodes in the same branch, and the lassos thus formed are such that the penalty of players that have a finite penalty threshold does not increase along them. We show that these completions of branches can be chosen of polynomial length. We then compress the core and expanded core so that they also have polynomial height.

The symbolic tree $\stree$ thus built is a subtree of $\tree$ (even if its unfolding $\widetilde\stree$ is not): in particular, as a corollary, if a player $j$ has no winning play in $\tree$, he does not have a winning play in $\stree$ neither. In particular, when we apply independently this proposition to all the trees of a forest $\forest$, to obtain a symbolic forest $\mathcal H$, this remark allows us to check that the new vector $\gamma^{\widetilde{\mathcal H}}$ has all its components not above the corresponding ones in $\gamma^{\forest}$ (if $\gamma^{\forest}_{i, v, \I} = 0$ then $\gamma^{\widetilde{\mathcal H}}_{i, v, \I}=0$). In particular, if the tree $\tree_{i, v, \I}$ of the forest $\forest$ is $(\gamma^{\forest},\playerSet\setminus \I)$-resistant, then the tree $\widetilde\stree_{i, v, \I}$ of the symbolic forest $\mathcal H$ is $(\gamma^{\widetilde{\mathcal H}},\playerSet\setminus \I)$-resistant. 

Finally, by combining this result with Remark~\ref{rem:gamma}, we obtain the following corollaries that allow us to obtain the PSPACE decision procedures: 
\begin{corollary}\label{cor:symbolic-NE}
Let $m \in (\N \cup \{ \infty \})^n$ be upper thresholds, and $M$ be the largest such upper threshold.
     The following assertions are equivalent:
    \begin{enumerate}
        \item There exists a permissive NE $\multiStrat$ in $\initGame$ such that:
    
        (a) $\multiStrat$ is strongly winning w.r.t. $\winningPlayers$;
        (b) for all $i \in \playerSet$, $\mainPenalty_i(\multiStrat,\initVertex) \leq m_i$.

    \item There exists a symbolic tree $\widetilde\tree$  in $\initGame$ of height polynomial in the number of players and vertices of $\game$ and in $M$, such that 
    
        (a) $\widetilde\tree$ is resistant to internal deviations, and $(\gamma^{\game}, \playerSet)$-resistant, with $\gamma^{\game}$ defined in Remark~\ref{rem:gamma}; and
        (b) 
        for all $\rho \in \widetilde\tree^\infty$ and $i \in \winningPlayers$, $\gain_i(\rho) = 1$;
        and (c) for all $i \in \playerSet$, $\Penalty_i(\widetilde\tree) \leq m_i$.
    
    \end{enumerate}
These assertions are still equivalent by replacing 1(a) by ``$\multiStrat$ is weakly winning w.r.t.~$\winningPlayers$'' and 2(b) by ``there exists $\rho \in \widetilde\tree^\infty$ such that for all $i \in \winningPlayers$, $\gain_i(\rho) = 1$''.
\end{corollary}

\begin{corollary}\label{cor:symbolic-SPE}
Let $m \in (\N \cup \{ \infty \})^n$ and $r \in (\N \cup \{ \infty \})^n$ be upper thresholds, and $M$ be the largest such upper threshold.
     The following assertions are equivalent:
    \begin{enumerate}
        \item There exists a permissive SPE $\multiStrat$ in $\initGame$ such that:
    
        (a) $\multiStrat$ is strongly winning w.r.t. $\winningPlayers$;
        and (b) for all $i \in \playerSet$, $\mainPenalty_i(\multiStrat,\initVertex) \leq m_i$ and $\retaliationPenalty_i(\multiStrat,\initVertex) \leq r_i$.

    \item There exists a symbolic forest $\forest$ in $\initGame$, where each symbolic tree has a height polynomial in the number of players and vertices of $\game$ and in $M$, such that 
      (a) each tree $\widetilde\tree_{i, v, \I}$ is $(\playerSet\setminus \I)$-resistant to internal deviations, and $(\gamma^{\forest}, \playerSet)$-resistant, with $\gamma^{\forest}$ defined in Remark~\ref{rem:gamma};
        (b) for all $\rho \in \widetilde\tree_{0,\initVertex,\initVisit}^\infty$ and $i \in \winningPlayers$, $\gain_i(\rho) = 1$;
        and (c) for all $i \in \playerSet$, $\mainPenalty_i(\widetilde\forest) \leq m_i$ and $\retaliationPenalty_i(\widetilde\forest) \leq r_i$.
    
    \end{enumerate}
    These assertions are still equivalent by replacing 1(a) by ``$\multiStrat$ is weakly winning w.r.t.~$\winningPlayers$'' and 2(b) by ``there exists $\rho \in \widetilde\tree^\infty_{0,\initVertex,\initVisit}$ such that for all $i \in \winningPlayers$, $\gain_i(\rho) = 1$''.
\end{corollary}

\subsection{Decision problems over permissive Nash equilibria}
\label{section:NEcomput}

For permissive NEs, it makes little sense to take into consideration the retaliation penalties, since the punishment after a deviation should definitely make the deviator lose whatever the penalty from now on. We thus obtain the following decision result: 
\begin{theorem}
    The constrained penalty problem, the weakly winning with constrained penalty problem and the strongly winning with constrained penalty problem, all with infinite (and thus no) constraints on retaliation penalties and for NEs are decidable in $\mathrm{PSPACE}$ (when the penalty bounds are encoded in unary). 
\end{theorem}
\begin{proof}
    We build upon Corollary~\ref{cor:symbolic-NE}, looking for a finite symbolic tree with the corresponding properties. 
    We first explain how to solve the constrained penalty problem, and explain afterwards the adaptation for the two other problems. The idea is to use an alternating polynomial time Turing machine (since $\mathrm{AP} = \mathrm{PSPACE}$ \cite{AP}) to guess a symbolic tree, checking the various constraints over it by using branch per branch. We describe the construction by supposing that the states of the Turing machine are split between existential states (where the machine accepts if at least one execution accepts) and universal states (where the machine accepts if all the executions accept). Existential states thus allow us to non-deterministically guess the finite symbolic tree node after node. We use a polynomial counter to keep track of the polynomially bounded height of the tree: if the counter goes over the polynomial bound, the execution of the alternating machine fails. At each node, existential states guess non-deterministically the set of successors on the working tape.

    Universal states allow us to check several pieces of information on the guessed symbolic tree: the resistance to internal deviations, the constraint on the penalty for each player, and the $(\gamma^{\game}, \playerSet)$-resistance, with $\gamma^{\game}$ as in Remark~\ref{rem:gamma}. Notice that this vector has exponential size, but the index $\I$ in a triple $(i, v, \I)$ is useless (apart from knowing if $i\in \I$), and can thus be ignored: moreover, this set $\I$ will be maintained along the execution of the algorithm. 
    This vector can thus be precomputed in (deterministic) polynomial time by determining, for each player, their set of winning vertices (against the coalition of the other players)~\cite{lncs2500}.
    
    The various checks can be performed branch per branch by keeping some pieces of information in memory, not only for the current node of the symbolic tree, but also for the whole current branch (this remains in polynomial space). Universal states are thus used to perform the checks on all the branches of the guessed tree.
    \begin{itemize}
        \item \emph{Checking the penalty for player~$i$.} If we have to check that the main penalty of player~$i$ is bounded by a threshold $m_i$ (\IE that the penalty of player~$i$ over each branch is bounded by $m_i$), we keep in memory the current penalty, forbidding for it to go above~$m_i$.
        
        \item \emph{Checking the resistance to internal deviations and $(\gamma^{\game},\playerSet)$-resistance.} At each node of the guessed tree, if the existential states guessed at least two successors, or depending on the vector $\gamma^{\game}$ (for a vertex $v$ where $\gamma^{\game}$ has value $1$, and that has not been chosen among the set of successors), we must remember constraints on the successors: either (\emph{a}) all plays in their subtrees must be winning for a certain player~$i$, or (\emph{b}) none.
        We could add neither constraint (\emph{a}) nor (\emph{b}) for a certain player (if only one successor has been chosen, and the $\gamma^{\game}$ value of all the other successors is 0). In the case where only the resistance to internal deviation applies (if at least two successors have been chosen, and the $\gamma^{\game}$ value of all the other successors is 0), the choice of constraint (\emph{a}) or (\emph{b}) is guessed non-deterministically.
        These constraints are kept all along the guessed branch except if a vertex of the target set of player~$i$ is visited; in this case the constraint (\emph{a}) is released.
        Moreover, the constraint (\emph b) for a player $i$ forbids to select a successor in the future where player $i$ visits one of his target vertices.
        \item \emph{The end of the branches.} The existential states decide when to stop the branch of the symbolic tree (before the counter runs out of the polynomial bound). Notice that the branch cannot stop if one of the type (\emph{a}) constraints is not released. Then existential states provide the set of successors taken in the ancestors so that for players that have a finite upper threshold on their penalty, ancestors must have the same current penalty as the leaf (to ensure that their penalty does not raise to $+\infty$ in the long run).
        
    \end{itemize}
    For the strongly winning variants, universal states also check the constraint that every player of $\winningPlayers$ must win at the end of each branch.    
    For the weakly winning variant, the existential states are also used to propose a branch where all players of $\winningPlayers$ will win. The universal states moreover check whether this condition is fulfilled for this particular branch.
\end{proof}

\subsection{Decision problems over permissive subgame perfect equilibria}
\label{section:SPEcomput}

\begin{theorem}
    The constrained penalty problem, the weakly winning with constrained penalty problem and the strongly winning with constrained penalty problem for SPEs are decidable in $\mathrm{PSPACE}$ (when the penalty bounds are encoded in unary). 
\end{theorem}
\begin{proof}
    The proof is the same as for NEs, instead of the fact that we use Corollary~\ref{cor:symbolic-SPE}, with a vector $\gamma^{\forest}$ that is partially guessed non-deterministically when it is needed. 
    When the existential states extend a branch of the tree $\tree$ from a vertex of player~$i$, the universal states does not only explore the chosen successors (with constraints (\emph{a}) or (\emph b) as in the previous proof), but now also explores the other vertices $u$ by starting a fresh exploration of another tree $\tree_{i, u, I}$ of the forest. Existential states also non-deterministically guess if player $i$ is weakly winning in $\tree_{i, u, I}$. If so, this gives new constraints (\emph a) in the tree $\tree$. The guessed weakly winning constraints are then checked in the fresh exploration: if player $i$ must be weakly winning, this is a constraint of the same type as a weakly winning constraint in the ``main'' tree; if player $i$ must not be weakly winning, this is a constraint of type (\emph b) (none of the play must be winning for player $i$) that we deal as before.

    Main penalties are checked as before. 
    For the retaliation penalties, for each player, we check that the total penalty of all new symbolic trees $\tree_{i, u, \I}$ is below the given upper threshold. 
    To ensure polynomial time termination, we maintain a polynomial counter, and the set of trees (more precisely, the set of triples $(i, u, I)$ used to index the trees of the forest) we jumped in so far. The polynomial counter again takes care of the depth of the branch we explore in the current tree (we reset this counter when we jump from a tree to another one). The set of trees we jumped in so far is maintained to forbid several explorations of the same tree of the forest. 
    As for NEs, the exploration is losing if the depth of the current branch is longer than the polynomial bound. The cardinal of the set of triples $(i, u, I)$ we must maintain is also polynomial (bounded by $|\playerSet|\times|V|\times |\playerSet|$, even though there are exponentially many trees in a forest), since the subset $\I$ of winning players does not decrease along the jumps from a tree to the next one. This also implies that the total length of the executions of the Turing machine is indeed polynomial.

    Notice that weakly and strongly winning conditions have only to be checked on the ``main'' tree as for permissive~NEs.
\end{proof}

\section{Conclusion}

We studied the permissiveness in Nash, and subgame perfect equilibria over multiplayer reachability games. We showed that several associated problems are decidable in PSPACE: they ask for the existence of such equilibria with various constraints, both on the set of players who reach their target set, and on the penalties that allow us to compare the permissiveness of two equilibria. The polynomial space depends on the size of the game, and the largest upper threshold on the penalties. We were not able to decrease the space dependency to be only polynomial in the logarithm of the penalty thresholds: we leave for future work to investigate if this is possible, or if there is a matching lower bound on~complexity. 

As other ideas for future works, we would like to extend our study to other objectives than reachability, like more general $\omega$-regular objectives (\emph{e.g.,}~parity games), but also weighted games like mean-payoff games, discounted-payoff games, or shortest-path games (where the reachability objective is combined with an objective to reach the target with the smallest possible total weight). An even more challenging problem is to extend this study to the setting of timed games, where the permissiveness is not only on the choice of edges, but also on the choice of delays spent in a given vertex. Work along these lines has been carried out on timed automata and two-player timed games~\cite{BouyerFM15,ClementJMM20}.

%\bibliography{biblio}

\begin{thebibliography}{}

\end{thebibliography}


\begin{thebibliography}{10}

    \bibitem{AnandNayakSchmuck}
    Ashwani Anand, Satya~Prakash Nayak, and Anne-Kathrin Schmuck.
    \newblock Synthesizing permissive winning strategy templates for parity games.
    \newblock In {\em {CAV} 2023}, volume 13964 of {\em LNCS}, pages 436--458. Springer, 2023.
    \newblock \href {https://doi.org/10.1007/978-3-031-37706-8_22} {\path{doi:10.1007/978-3-031-37706-8_22}}.
    
    \bibitem{BJW02}
    Julien Bernet, David Janin, and Igor Walukiewicz.
    \newblock Permissive strategies: from parity games to safety games.
    \newblock {\em {RAIRO} Theor. Informatics Appl.}, 36(3):261--275, 2002.
    \newblock \href {https://doi.org/10.1051/ITA:2002013} {\path{doi:10.1051/ITA:2002013}}.
    
    \bibitem{BDMR09}
    Patricia Bouyer, Marie Duflot, Nicolas Markey, and Gabriel Renault.
    \newblock Measuring permissivity in finite games.
    \newblock In {\em {CONCUR} 2009}, volume 5710 of {\em LNCS}, pages 196--210. Springer, 2009.
    \newblock \href {https://doi.org/10.1007/978-3-642-04081-8\_14} {\path{doi:10.1007/978-3-642-04081-8\_14}}.
    
    \bibitem{BouyerFM15}
    Patricia Bouyer, Erwin Fang, and Nicolas Markey.
    \newblock Permissive strategies in timed automata and games.
    \newblock {\em Electron. Commun. Eur. Assoc. Softw. Sci. Technol.}, 72, 2015.
    \newblock \href {https://doi.org/10.14279/TUJ.ECEASST.72.1015} {\path{doi:10.14279/TUJ.ECEASST.72.1015}}.
    
    \bibitem{BouyerMOU11}
    Patricia Bouyer, Nicolas Markey, J{\"{o}}rg Olschewski, and Michael Ummels.
    \newblock Measuring permissiveness in parity games: Mean-payoff parity games revisited.
    \newblock In {\em {ATVA} 2011}, volume 6996 of {\em LNCS}, pages 135--149. Springer, 2011.
    \newblock \href {https://doi.org/10.1007/978-3-642-24372-1\_11} {\path{doi:10.1007/978-3-642-24372-1\_11}}.
    
    \bibitem{brice2022}
    L\'{e}onard Brice, Jean-Fran\c{c}ois Raskin, and Marie van~den Bogaard.
    \newblock {On the Complexity of SPEs in Parity Games}.
    \newblock In {\em {CSL} 2022}, volume 216 of {\em LIPIcs}, pages 10:1--10:17. Schloss Dagstuhl -- Leibniz-Zentrum f{\"u}r Informatik, 2022.
    \newblock \href {https://doi.org/10.4230/LIPIcs.CSL.2022.10} {\path{doi:10.4230/LIPIcs.CSL.2022.10}}.
    
    \bibitem{briceMeanPayoff}
    L\'{e}onard Brice, Jean-Fran\c{c}ois Raskin, and Marie van~den Bogaard.
    \newblock {The Complexity of SPEs in Mean-Payoff Games}.
    \newblock In {\em {ICALP} 2022}, volume 229 of {\em LIPIcs}, pages 116:1--116:20. Schloss Dagstuhl -- Leibniz-Zentrum f{\"u}r Informatik, 2022.
    \newblock \href {https://doi.org/10.4230/LIPIcs.ICALP.2022.116} {\path{doi:10.4230/LIPIcs.ICALP.2022.116}}.
    
    \bibitem{Brice}
    Léonard Brice, Jean-François Raskin, and Marie van~den Bogaard.
    \newblock Subgame-perfect equilibria in mean-payoff games.
    \newblock {\em Logical Methods in Computer Science}, 19, 2023.
    \newblock \href {https://doi.org/10.46298/LMCS-19(4:6)2023} {\path{doi:10.46298/LMCS-19(4:6)2023}}.
    
    \bibitem{weakSPEs}
    Thomas Brihaye, V{\'{e}}ronique Bruy{\`{e}}re, Aline Goeminne, and Jean{-}Fran{\c{c}}ois Raskin.
    \newblock Constrained existence problem for weak subgame perfect equilibria with {\(\omega\)}-regular boolean objectives.
    \newblock In {\em {GandALF} 2018}, volume 277 of {\em {EPTCS}}, pages 16--29, 2018.
    \newblock \href {https://doi.org/10.4204/EPTCS.277.2} {\path{doi:10.4204/EPTCS.277.2}}.
    
    \bibitem{BrihayeBGRB19}
    Thomas Brihaye, V{\'{e}}ronique Bruy{\`{e}}re, Aline Goeminne, Jean{-}Fran{\c{c}}ois Raskin, and Marie van~den Bogaard.
    \newblock The complexity of subgame perfect equilibria in quantitative reachability games.
    \newblock In {\em {CONCUR} 2019}, volume 140 of {\em LIPIcs}, pages 13:1--13:16. Schloss Dagstuhl - Leibniz-Zentrum f{\"{u}}r Informatik, 2019.
    \newblock \href {https://doi.org/10.4230/LIPICS.CONCUR.2019.13} {\path{doi:10.4230/LIPICS.CONCUR.2019.13}}.
    
    \bibitem{BrihayeBGT19}
    Thomas Brihaye, V{\'{e}}ronique Bruy{\`{e}}re, Aline Goeminne, and Nathan Thomasset.
    \newblock On relevant equilibria in reachability games.
    \newblock In {\em {RP} 2019}, volume 11674 of {\em LNCS}, pages 48--62. Springer, 2019.
    \newblock \href {https://doi.org/10.1007/978-3-030-30806-3\_5} {\path{doi:10.1007/978-3-030-30806-3\_5}}.
    
    \bibitem{BrihayeBMR15}
    Thomas Brihaye, V{\'{e}}ronique Bruy{\`{e}}re, No{\'{e}}mie Meunier, and Jean{-}Fran{\c{c}}ois Raskin.
    \newblock Weak subgame perfect equilibria and their application to quantitative reachability.
    \newblock In {\em {CSL} 2015}, volume~41 of {\em LIPIcs}, pages 504--518. Schloss Dagstuhl - Leibniz-Zentrum f{\"{u}}r Informatik, 2015.
    \newblock \href {https://doi.org/10.4230/LIPICS.CSL.2015.504} {\path{doi:10.4230/LIPICS.CSL.2015.504}}.
    
    \bibitem{AP}
    Ashok~K. Chandra, Dexter~C. Kozen, and Larry~J. Stockmeyer.
    \newblock Alternation.
    \newblock {\em Journal of the ACM}, 28(1):114–133, 1981.
    
    \bibitem{ClementJMM20}
    Emily Clement, Thierry J{\'{e}}ron, Nicolas Markey, and David Mentr{\'{e}}.
    \newblock Computing maximally-permissive strategies in acyclic timed automata.
    \newblock In {\em {FORMATS} 2020}, volume 12288 of {\em LNCS}, pages 111--126. Springer, 2020.
    \newblock \href {https://doi.org/10.1007/978-3-030-57628-8\_7} {\path{doi:10.1007/978-3-030-57628-8\_7}}.
    
    \bibitem{lncs2500}
    Erich Gr{\"{a}}del, Wolfgang Thomas, and Thomas Wilke, editors.
    \newblock {\em Automata, Logics, and Infinite Games: {A} Guide to Current Research [outcome of a Dagstuhl seminar, February 2001]}, volume 2500 of {\em LNCS}. Springer, 2002.
    \newblock \href {https://doi.org/10.1007/3-540-36387-4} {\path{doi:10.1007/3-540-36387-4}}.
    
    \bibitem{NayakS24}
    Satya~Prakash Nayak and Anne{-}Kathrin Schmuck.
    \newblock Most general winning secure equilibria synthesis in graph games.
    \newblock In {\em {TACAS} 2024}, volume 14572 of {\em LNCS}, pages 173--193. Springer, 2024.
    \newblock \href {https://doi.org/10.1007/978-3-031-57256-2\_9} {\path{doi:10.1007/978-3-031-57256-2\_9}}.
    
    \bibitem{Ummels06}
    Michael Ummels.
    \newblock Rational behaviour and strategy construction in infinite multiplayer games.
    \newblock In {\em {FSTTCS} 2006}, volume 4337 of {\em LNCS}, pages 212--223. Springer, 2006.
    \newblock \href {https://doi.org/10.1007/11944836\_21} {\path{doi:10.1007/11944836\_21}}.
    
    \end{thebibliography}

\appendix

\section{Proofs of Section~\ref{section:charac}: \nameref{section:charac}}

\subsection{\nameref{section:NEcharac}}

\thmNECharac*

\begin{proof}~
    \begin{itemize}
        \item ($1 \Rightarrow 2$) Let us assume that there exists a permissive NE $\multiStrat$ in $\initGame$ such that $\multiHistory{\multiStrat}{\initVertex} = \tree$. 
        \begin{itemize}
            \item First, let us assume by contradiction that the tree $\tree$ is not resistant to internal deviations. Therefore, there exists $hv \in \tree$ with $v \in V_i$ (for some $i\in \playerSet$) such that $|\{hvv'\in \tree\mid v' \in V\}| \geq 2$, as well as two plays $\rho, \rho'\in \tree^\infty_{\restriction hv}$ such that $\gain_i(h\rho)=0$ and $\gain_i(h\rho')=1$. If $\rho$ and $\rho'$ start by the same vertex, by using the fact that $hv$ has at least two children in $\tree$, we can find another play $\rho''$ in $\tree^\infty_{\restriction hv}$ that starts by jumping in child different from the ones taken in $\rho$ and $\rho'$. In particular, the play $h\rho''$ has a gain for player $i$ different from either $\rho$ or $\rho'$. If $\rho$ and $\rho'$ jump in different vertices, we directly get the same conclusion. 

        From now, we thus suppose that $\rho$ and $\rho'$ do not jump in the same vertex on their first step. 
        We show that player $i$ has a profitable deviation in $\multiStrat$, contradicting the fact that this is a permissive NE. Indeed, let $w$ and $w'$ be the (different) second vertices in $\rho$ and $\rho'$ respectively.
        Consider a profile $\sigma$ of strategies consistent with $\multiStrat$ such that the outcome from $\initVertex$ is $h\rho$ and the outcome from history $hvw'$ is $\rho'_{\geq 1}$. Consider the strategy $\sigma'_i$ of player $i$ that plays according to $\sigma$ unless in history $hv$ where he chooses successor $w'$ instead of $w$. Then, we have $\gain_i(\outcome{\sigma}{v_0})=0$ and $\gain_i(\outcome{\sigma'_i,\sigma_{-i}}{v_0})=1$ which contradicts the hypothesis that $\sigma$ should be an NE (since it is consistent with the permissive NE $\multiStrat$). Thus, the tree $\tree$ is resistant to internal deviations.

        \item Then, let us assume by contradiction that the tree $\tree$ is not resistant to external deviations. Thus, there exists $hu \in \tree$ with $u \in V_i$ for some $i\in \playerSet\setminus \visit(hu)$, and $u'\in \successor(u)$ with $huu'\notin\tree$ such that player $i$ has a winning strategy from $u'$, 
        but there is a play $\rho \in \tree^\infty_{\restriction hu}$ such  that $\gain_i(\rho)=0$. 
        Consider a profile $\sigma$ of strategies consistent with $\multiStrat$ such that the outcome from $\initVertex$ is $h\rho$. Consider the strategy $\sigma'_i$ of player $i$ that plays according to $\sigma$ unless in history $hu$ where it goes to $u'$ before following a winning strategy of player~$i$ from now on. Then, we have $\gain_i(\outcome{\sigma}{v_0})=0$ and $\gain_i(\outcome{\sigma'_i,\sigma_{-i}}{v_0})=1$ which again contradicts the hypothesis that $\sigma$ should be an NE. Thus, $\tree$ is resistant to external deviations, and is thus a good tree.
        \end{itemize}
        
        \item ($2 \Rightarrow 1$) We suppose that $\tree$ is good, and build a permissive NE $\multiStrat$ by following the decisions in the tree $\tree$, and for histories $huh'$ such that $h\in \tree$ with the last vertex of $h$ belonging to player $i$ but $hu\notin \tree$, letting all players $j\neq i$ play a coalition strategy against player $i$ maximizing their gain (thus making player $i$ lose if possible, by determinacy of two-player reachability games). We show that $\multiStrat$ is a permissive NE. To do so, consider a profile of strategies $\sigma$ consistent with $\multiStrat$. Suppose by contradiction that it contains a profitable deviation for player $i$, \IE that there exists $\sigma'_i$ such that $\gain_i(\outcome{\sigma}{v_0})=0$ and $\gain_i(\outcome{\sigma'_i,\sigma_{-i}}{v_0})=1$.
        Let $h$ be the first history of $\outcome{\sigma}{v_0}$ where $\sigma_i$ and $\sigma'_i$ take different actions, that we call $v$ and $v'$. If $hv'\in \tree$, since $\tree$ is resistant to internal deviations, we know that $\outcome{\sigma'_i,\sigma_{-i}}{v_0}$ cannot be a play in $\tree^\infty$. Thus, there exists a longer history $h'$ of $\outcome{\sigma'_i,\sigma_{-i}}{v_0}$ with $\last(h')\in V_i$ where $\sigma'_i(h')$ plays outside $\tree$. If $hv'\notin \tree$, we directly obtain such a history $h'$. Notice that $i\notin \visit(h')$ since $\gain_i(\outcome{\sigma}{v_0})=0$. However, this means that player~$i$ has found a way to win after history~$h'$ thus just went outside the tree. In particular,
        player~$i$ has a winning strategy against the coalition of the other players from history $h'$.  Moreover, there exists a play inside the tree from $h'$ such that player~$i$ should win (by resistance to external deviations). Nonetheless this winning play contradicts the resistance of internal deviations because $\gain_i(\outcome{\sigma}{\initVertex}) = 0$.\qedhere
        
        %However, this means that player $i$ has found a way to win after history $h'$ thus just went outside the tree although he was losing in at most  one play of the tree . We thus know that player $i$ has a winning strategy against the coalition of the other players from history $h'$ which would imply, by resistance to external deviations, that all plays in $\tree$ from $h'$ is also winning for~$i$. This contradicts the existence of the play $\outcome{\sigma}{v_0}$.\qedhere
        \end{itemize}
\end{proof}

\subsection{\nameref{section:SPEcharac}}

\thmSPECharac*

The proof of Theorem~\ref{thm:SPECharac} relies on the fact that in reachability games the notion of SPEs is equivalent to a weaker notion of equilibrium called \emph{weak subgame perfect equilibrium}.
Given a strategy $\sigma_i$ of player~$i$, a strategy $\sigma'_i$ of player~$i$ such that $\sigma_i \neq \sigma'_i$ is called a \emph{deviating strategy} from $\sigma_i$. 
Furthermore, if $\sigma'_i$ differs from $\sigma_i$ on a finite number  of histories (resp.~only on $\initVertex$), $\sigma'_i$ is \emph{finitely deviating} (resp.~\emph{one-shot deviating}) from $\sigma_i$. 

When considering NEs and SPEs the set of possible deviating strategies for a player is not restricted. One can also consider that a player may only use a finitely (resp.~one-shot) deviating strategy. This restriction leads to the notion of (very) weak NEs and SPEs. Formally, a strategy profile $\sigma$ is a weak NE (resp.~very weak NE) in $\initGame$ if for all $i \in \playerSet$ and all $\sigma'_i$ which is finitely (resp.~one-shot) deviating from $\sigma_i$, $\gain_i(\outcome{\sigma}{v_0}) \geq \gain_i(\outcome{\sigma'_i,\sigma_{-i}}{v_0})$.
In the same way, a strategy profile $\sigma$ is a weak (resp.~very weak) SPE in $\initGame$ if for all $i \in \playerSet$, for all $hv \in \hist_i(\initVertex)$, for all $\sigma'_i$ strategy of player~$i$ in $\initSubgame{h}{v}$, $\sigma_{\restriction h}$ is a weak (resp.~very weak) NE in $\initSubgame{h}{v}$. 
Notice that a strategy profile is a weak SPE if and only if it is a very weak SPE~\cite{BrihayeBMR15}. Even more, in reachability games notions of (very) weak SPEs and SPEs coincide.

\begin{proposition}[\cite{BrihayeBMR15}]
    A strategy profile $\sigma$ is a (very) weak SPE in the reachability game $\initGame$ if and only if it is an SPE in $\initGame$.
\end{proposition}

In order to ease the proof of Theorem~\ref{thm:SPECharac}, we first prove an intermediate result about permissive SPEs.

\begin{restatable}{proposition}{intermediateSPEcharac}
\label{prop:intermediateSPEcharac}
    Let $\multiStrat$ be a permissive SPE in $\initGame$. Then 
    \begin{enumerate}
        \item for all $i \in \playerSet$ and for all $hv \in \hist_i(\initVertex)$, if $|\multiStrat_i(hv)| \geq 2$, then for all $\rho, \rho' \in \multiOutcome{\multiStrat_{\restriction h}}{v}$, $\gain_i(h\rho) = \gain_i(h\rho')$;
        \item for all $i \in \playerSet$ and $hv \in \hist_i(\initVertex)$ such that there exists $u' \in \successor(v) \setminus \multiStrat_i(hv)$, if there exists $\rho' \in \multiOutcome{\multiStrat_{\restriction hv}}{u'}$ such that $\gain_i(hv\rho') = 1$, then for all $\rho \in \multiOutcome{\multiStrat_{\restriction h}}{v}$, $\gain_i(h\rho) = 1$.
    \end{enumerate}
\end{restatable}
\begin{proof}
    %Let $\multiStrat$ be a permissive SPE in $\initGame$.
    \begin{enumerate}
        %Interal deviation
        \item Let $i \in \playerSet$ and $hv \in \hist_i(\initVertex)$ such that $|\multiStrat_i(hv)| \geq 2$. We assume that there exist $\rho, \rho' \in \multiOutcome{\multiStrat_{\restriction h}}{v}$ such that $\gain_i(h\rho) = 0$ and $\gain_i(h\rho') =1$. 
        %Notice that, that means that $i \not \in \visit(hv)$.
        Let $u$ and $u'$ be the first vertices of $\rho$ and $\rho'$ respectively.  
        \begin{enumerate}
           \item  If $u \neq u'$ (possible since $|\multiStrat_i(hv)| \geq 2$), we consider a strategy $\sigma$ in $\initGame$ such that $\sigma \lesssim \multiStrat$ with $\outcome{\sigma_{\restriction h}}{v} = \rho$ and $\outcome{\sigma_{\restriction hv}}{u'} = \rho'_{\geq 1}$. We claim that $\sigma$ is not a very weak SPE (and so not an SPE) in $\initGame$. Indeed, in the subgame $\initSubgame{h}{v}$, the one-shot deviating strategy $\sigma'_i$ from $\sigma_{i \restriction h}$ such that $\sigma'_i(v) = u'$ satisfies 
           \begin{align*}\gain_{i \restriction h}(\outcome{\sigma'_i, \sigma_{-i \restriction h}}{v}) &= \gain_{i \restriction h}(v\outcome{\sigma_{i \restriction hv},\sigma_{-i \restriction hv}}{u'})= \gain_{i\restriction h}(\rho') = 1
           \end{align*}
           and
           $\gain_{i\restriction h}(\outcome{\sigma_{\restriction h}}{v}) = \gain_{i \restriction h} (\rho) = 0$.

           \item If $u' = u$, since $|\multiStrat_i(hv)| \geq 2$, there exists $u'' \neq u$ such that $u'' \in \multiStrat_i(hv)$ and there exists $\rho'' \in \multiOutcome{\multiStrat_{\restriction h}}{v}$ such that $u''$ if the first vertex of $\rho''$. Thus, either $\gain_i(\rho'') = 1$ and we repeat the previous argument with $\rho$ and $\rho''$ or $\gain_i(\rho'') = 0$ and we repeat the argument with $\rho'$ and $\rho''$.  
        \end{enumerate}

        % External deviations

        \item Let $i \in \playerSet$ and $hv \in \hist_i(\initVertex)$ such that there exist $u' \in \successor(v) \setminus \multiStrat_i(hv)$ and $\rho' \in \multiOutcome{\multiStrat_{\restriction hv}}{u'}$ with $\gain_i(hv\rho') =1$. If there exists $\rho \in \multiOutcome{\multiStrat_{\restriction h}}{v}$ such that $\gain_i(h\rho) = 0$, then any strategy profile $\sigma \lesssim \multiStrat$ such that  $\outcome{\sigma_{\restriction h}}{v} = \rho$ and $\outcome{\sigma_{\restriction hv}}{u'} = \rho'$ cannot be a very weak SPE. Indeed, let us consider the subgame $\initSubgame{h}{v}$ and the one-shot deviating strategy $\sigma'_i$ from $\sigma_{i \restriction h}$ such that $\sigma'_i(v) = u'$: this is a true deviation since $u' \notin \multiStrat_i(hv)$ and $\sigma_i(hv) \in \multiStrat_i(hv)$ by definition. Then,
           \begin{align*}\gain_{i \restriction h}(\outcome{\sigma'_i, \sigma_{-i \restriction h}}{v}) &= \gain_{i \restriction h}(v\outcome{\sigma_{i \restriction hv},\sigma_{-i \restriction hv}}{u'})
           = \gain_{i\restriction h}(\rho') = 1
           \end{align*}
           and
           $\gain_{i\restriction h}(\outcome{\sigma_{\restriction h}}{v}) = \gain_{i \restriction h} (\rho) = 0$.    \qedhere 
    \end{enumerate}
\end{proof}

%We now prove Theorem~\ref{thm:SPECharac}.
\begin{proof}[Proof of Theorem~\ref{thm:SPECharac}]
% Let $m \in (\N \cup \{ \infty \})^n$ and $r \in (\N \cup \{ \infty \})^n$ be upper thresholds. Let $\tree^*$ be a tree rooted at $\initVertex$ and $\winningPlayers \subseteq \playerSet$ be a set of players.
~

\noindent \underline{$(1 \Rightarrow 2)$} Let us assume that there exists a permissive SPE $\multiStrat$ in $\initGame$ such that properties 1a, 1b, and 1c hold.
    % \emph{(i)}~$\multiHistory{\multiStrat}{\initVertex} = \tree^*$;
    %     \emph{(ii)}~$\multiStrat$ is strongly winning w.r.t. $\winningPlayers$; and \emph{(iii)} for all $i \in \playerSet$, $\mainPenalty_i(\multiStrat,\initVertex) \leq m_i$ and $\retaliationPenalty_i(\multiStrat,\initVertex) \leq r_i$.
        %
       % We first build a forest $\forest$ from $\multiStrat$. For all $v \in v$, we let $\tree^C_v$ be the complete tree from $v$. For each $(i,v, \I) \in \mathcal{I}$, we define $$\mathcal{H}(i,v,\I) = \{hv \in \hist(\initVertex) \setminus \multiHistory{\multiStrat}{\initVertex} \, \wedge \, \last(h) \in V_i \, \wedge \, \I = \visit(hv) \}$$ and 
        %$$\mathcal{O}(i,v, \I) = 
        %\begin{cases} 
        %\{ \multiHistory{\multiStrat_{\restriction h}}{v} \mid hv \in \mathcal{H}(i,v,\I) \} &  \text{ if } \mathcal{H}(i,v,\I) \neq \emptyset \\
        %\{ \tree^C_v \} &  \text{ otherwise} \end{cases}.$$
%
         We first build a forest $\forest$ from $\multiStrat$. For each $(i,v, \I) \in \mathcal{I}$, we let
        \[\mathcal{O}(i,v, \I) =  \{ \multiHistory{\multiStrat_{\restriction h}}{v} \mid hv \in \hist(\initVertex)  \, \wedge \, \last(h) \in V_i \, \wedge \, \I = \visit(hv) \}\]
        We choose $\tree_{0,\initVertex, \initVisit} = \multiHistory{\multiStrat}{\initVertex} = \tree^*$ and for each  $(i,v, \I) \in \mathcal{I}$ such that $(i,v,\I) \neq (0, \initVertex, \initVisit)$, we choose $\tree_{i,v,\I} \in \mathcal{O}(i,v,\I)$ such that
        \begin{equation}
            \max\{ \gain_i(\rho) \mid \rho \in \tree^\infty_{i,v, \I} \} = \min_{\tree \in \mathcal{O}(i,v,\I)} \max \{ \gain_i(\rho) \mid \rho \in \infTree\}. \label{eq:witness1}
        \end{equation}
We have to prove that $\forest$ is good. Let $\tree_{i,v,\I} \in \forest$.
\begin{itemize}
    \item \textbf{(Resistance to internal deviations)}  
    Let us prove that $\tree_{i,v,\I}$ is $(\playerSet \setminus \I)$-resistant to internal deviations.
    Let $j \in (\playerSet \setminus \I)$ and  $gu \in \tree_{i,v,\I}$ be such that $u \in V_j$ and $|\{guu'\in \tree_{i,v,\I} \mid u' \in V\}| \geq 2$. We have to prove that for all $\rho,\rho' \in \tree^\infty_{i,v,I \restriction gu}, \gain_j(g\rho) = \gain_j(g\rho')$.
    
    We have that $\tree^\infty_{i,v,\I} = \multiOutcome{\multiStrat_{\restriction h}}{v} $ for some $hv \in \hist(\initVertex)$ such that $\last(h) \in V_i$ and $\I = \visit(hv)$. Moreover, since $|\{guu'\in \tree_{i,v,\I} \mid u' \in V\}| \geq 2$, we have $|\multiStrat_j(hgu)| \geq 2$. By Proposition~\ref{prop:intermediateSPEcharac}, we have that for all $\rho, \rho' \in \multiOutcome{\multiStrat_{\restriction hg}}{u} =  \tree^\infty_{i,v,I \restriction gu}$, $\gain_j(hg\rho) = \gain_j(hg\rho')$. We conclude that $\gain_j(g\rho')= \gain_j(g\rho)$ because $j \not \in \I = \visit(hv)$.

    \item \textbf{(Resistance to constrained external deviations)} 
    Let us prove that $\tree_{i,v,\I}$ is resistant to constrained external deviations.  Let $gu \in  \tree_{i,v,\I}$ and $j \in \playerSet$ such that we have that $u \in V_j$, $j \not \in I \cup \visit(gu)$, and there exists $u' \in \successor(u)$ such that $guu' \not \in \tree_{i,v,\I}$. Moreover, let us assume that there exists $\rho' \in \tree^\infty_{j,u',\I'}$, where $\I' = \I \cup \visit(guu')$, such that $\gain_j(\rho') = 1$. Let $\rho \in \tree^\infty_{i,v,\I\restriction gu}$, we have to prove that $\gain_j(\rho) =1$.

    By construction of $\forest$, there exists $hv \in \hist(\initVertex)$ such that $\last(h) \in V_i$, $\visit(hv) = \I$ and $\multiOutcome{\multiStrat_{\restriction h}}{v} = \tree^\infty_{i,v,\I}$. 
    We also have that $u' \not \in \multiStrat_i(hgu)$ (as $guu' \not \in \tree_{i,v,\I}$) and $\multiHistory{\multiStrat_{\restriction hgu}}{u'} \in \mathcal{O}(j,u',\I')$.
    In particular,
    \begin{align*}
        \max \{ \gain_j(\pi) \mid \pi \in \multiOutcome{\multiStrat_{\restriction hgu}}{u'}\}&\geq  \min_{\tree \in \mathcal{O}(j,u',\I')} \max \{ \gain_j(\pi) \mid \pi \in \tree^\infty\} \\
        &= \max \{ \gain_j(\pi) \mid \pi \in \tree^\infty_{j,u',\I'}\}= \gain_j(\rho')=1 
    \end{align*}
    It follows that there exists $\pi' \in \multiOutcome{\multiStrat_{\restriction hgu}}{u'} $ such that $\gain_j(\pi') = 1$ and so $\gain_j(hgu\pi') = 1$.
   By Proposition~\ref{prop:intermediateSPEcharac},  we have that for all $\pi \in \multiOutcome{\multiStrat_{\restriction hg}}{u}$, $\gain_j(\pi) = 1$. To conclude, as $\rho \in \tree^\infty_{i,v\I \restriction gu} = \multiOutcome{\multiStrat_{\restriction hg}}{u}$, $\gain_j(\rho) = 1$.
    \end{itemize}

    We have proved that $\forest$ is a good forest with $\tree_{0,\initVertex,\initVisit} = \tree^*$. Since $\multiStrat$ is strongly winning w.r.t.~$\winningPlayers$, we have that for all $\sigma \lesssim \multiStrat$, for all $i \in \winningPlayers$, $\gain_i(\outcome{\sigma}{\initVertex}) = 1$. It follows that for all $\rho \in \multiOutcome{\multiStrat}{\initVertex} = \tree^\infty_{0,\initVertex,\initVisit}$ and for all $i \in \winningPlayers$, $\gain_i(\rho) =1$. The same kind of argument holds if we replace assertion~\ref{assert:strongWinStrat} by ``$\multiStrat$ is weakly winning w.r.t.~$\winningPlayers$'' and the assertion~\ref{assert:strongWinWitness} by ``there exists $\rho \in \tree^\infty_{i,\initVertex,\initVisit}$ such that for all $i \in \winningPlayers$, $\gain_i(\rho) = 1$''.

    We finally prove that for all $i \in \playerSet$, $\mainPenalty_i(\forest) \leq m_i$ and $\retaliationPenalty_i(\forest) \leq r_i$. For~$i \in \playerSet$,
    \begin{align*}
        \mainPenalty_i(\forest) 
        &= \penalAux_i(\tree_{0,\initVertex,\initVisit})
        = \sup\{\penalAux_i(\rho) \mid \rho \in \tree^\infty_{0,\initVertex,\initVisit} \}\\
        &= \sup\{\penal{i}{\multiStrat}{\rho} \mid \rho \in \multiOutcome{\multiStrat}{\initVertex}\}
        = \mainPenalty_i(\multiStrat,\initVertex) \leq m_i.
    \end{align*}
    and since $\displaystyle \{ \tree_{j,u,\J} \mid (j,u,\J) \in \Out \} \subseteq \{ \multiHistory{\multiStrat_{\restriction h}}{v} \mid hv \in \hist(\initVertex) \setminus \multiHistory{\multiStrat}{\initVertex} \}$,
    \begin{align*}
        \retaliationPenalty_i(\forest) 
        &= \sup_{\substack{ \tree_{j,u,\J} \in \forest \\ (j,u,\J) \in \Out }}  
        \penalAux_i(\tree_{j,u,\J})
        \leq \sup_{\substack{ \multiHistory{\multiStrat_{\restriction h}}{v} \\ hv \in \hist(\initVertex) \setminus \multiHistory{\multiStrat}{\initVertex} }} \penalAux_i(
        {\multiHistory{\multiStrat_{\restriction h}}{v}}) \\ 
        %&= \sup_{\substack{ \multiHistory{\multiStrat_{\restriction h}}{v} \\ hv \in \hist(\initVertex) \setminus \multiHistory{\multiStrat}{\initVertex} }} \sup \{ \penalAux_i(\rho) \mid \rho \in \multiOutcome{\multiStrat_{\restriction h}}{v} \} \\
        &= \sup_{ hv \in \hist(\initVertex) \setminus \multiHistory{\multiStrat}{\initVertex}} \sup \{ \penalAux_i(\rho) \mid \rho \in \multiOutcome{\multiStrat_{\restriction h}}{v} \}
        = \retaliationPenalty_i(\multiStrat,\initVertex) \leq r_i
    \end{align*}

% 2 => 1
\medskip 
\noindent \underline{$(2 \Rightarrow 1)$} Let us assume that there exists a good forest $\forest$ in $\initGame$ such that properties 2a, 2b, and 2c hold. 
% : \emph{(i)} $\tree_{i,\initVertex, \initVisit} = \tree^*$; \emph{(ii)} for all $\rho \in \tree^\infty_{0,\initVertex,\initVisit}$, for all $i \in \winningPlayers$, $\gain_i(\rho) = 1$; and \emph{(iii)} for all $i \in \playerSet$, $\mainPenalty_i(\forest) \leq m_i$ and $\retaliationPenalty_i(\forest) \leq r_i$.
%
    We start defining a multi-strategy $\multiStrat$ such that $\multiHistory{\multiStrat}{\initVertex} = \tree_{0,\initVertex,\initVisit} = \tree^*$. Then, we continue building $\multiStrat$ by induction on the length of histories $hv$. Let us assume that for all $hv \in \hist(\initVertex)$ such that $|hv|=k$, if $v \in V_i$, then $\multiStrat_i(hv)$ is already defined. 
Let $hvv' \in \hist(\initVertex)$ such that $|hvv'|= k+1$, and if $v \in V_i$ and $v' \in V_j$, $\multiStrat_i(hv)$ is defined but not yet $\multiStrat_j(hvv')$.
We then extend the definition of $\multiStrat$ such that $\multiHistory{\multiStrat_{\restriction hv}}{v'} = \tree_{i,v', \I'}$ where $\I' = \visit(hvv')$.

Let us prove that $\multiStrat$ is a permissive SPE in $\initGame$.
Let $\sigma = (\sigma_1, \ldots, \sigma_n)$ be a strategy profile such that for all $i \in \playerSet$, $ \sigma_i \lesssim  \multiStrat_i$.
Let us prove that $\sigma$ is a (very) weak SPE in $\initGame$ (and so an SPE).

Let $hv \in \hist(\initVertex)$ with $v \in V_i$ for some $i \in \playerSet$. 
Let $\sigma'_i$ be a one-shot deviating strategy from $\sigma_{i\restriction h}$ in $(\game_{\restriction h}, v)$. Let us assume that $\sigma'_i(v) = u' \neq  u = \sigma_{i \restriction h} (v)$ for some $u,u' \in \successor(v)$.

By construction of $\multiStrat$,
there exists $h'w \leq h$ such that $h' \in \hist_j(\initVertex)$ for some $j \in \playerSet$ 
and $\multiHistory{\multiStrat_{\restriction h'}}{w} = \tree_{j,w, J}$ where $J = \visit(h'w)$. Moreover, by writing $h$ as $h'g$ for some history $g \in \hist(w)$, we have that $g \outcome{\sigma_{\restriction h}}{v} \in \tree^{\infty}_{j,w,J}$. We have to prove that $$\gain_{i \restriction h}(\outcome{\sigma_{\restriction h}}{v}) \geq \gain_{i\restriction h}(\outcome{\sigma'_{i}, \sigma_{-i \restriction h}}{v}).$$

Let us first notice that if $i \in \visit(hv)$, then $\gain_{i \restriction h}(\outcome{\sigma_{\restriction h}}{v}) = \gain_{i\restriction h}(\outcome{\sigma'_{i}, \sigma_{-i \restriction h}}{v}) =1$. Thus, we consider that $i \not\in \visit(hv)$ and in particular $i \not \in J$.
We distinguish two cases, depending on whether $g\outcome{\sigma'_i, \sigma_{-i\restriction h}}{v}$ is also in $\tree^{\infty}_{j,w,J}$ or not.

\begin{itemize}
    \item If $g\outcome{\sigma'_i, \sigma_{-i\restriction h}}{v} \in \tree^{\infty}_{j,w,J}$, since $\tree^{\infty}_{j,w,J}$ is part of $\forest$, it is $(\playerSet \setminus J)$-resistant to internal deviations. Thus, as $gvu$, $gvu' \in \tree_{j,w,J}$ and $u\neq u'$, we have that $|\{ gvv'' \in \tree_{j,w,J}\mid v'' \in V\}| \geq 2$. It follows from the definition of being $(\playerSet\setminus J)$-resistant to internal deviations and the fact that $i \not \in J$ that for all $\rho, \rho' \in \tree^\infty_{j,w,J \restriction gv}$, $\gain_i(g\rho) = \gain_i(g\rho')$. In particular, $\gain_i(g\outcome{\sigma_{\restriction h}}{v})=\gain_i( g\outcome{\sigma'_i, \sigma_{-i\restriction h}}{v})$ and $\gain_{i \restriction h}(\outcome{\sigma_{\restriction h}}{v}) = \gain_i(h'g\outcome{\sigma_{\restriction h}}{v})=\gain_i( h'g\outcome{\sigma'_i, \sigma_{-i\restriction h}}{v}) = \gain_{i\restriction h}( \outcome{\sigma'_i, \sigma_{-i\restriction h}}{v})$.
    We can conclude that $\sigma'_i$ is not a profitable deviation in $(\game_{\restriction h},v)$.

    \item If $g\outcome{\sigma'_i, \sigma_{-i\restriction h}}{v} \not\in \tree^{\infty}_{j,w,J}$, then $\outcome{\sigma'_{i\restriction v}, \sigma_{-i\restriction hv}}{u'} \in \tree^\infty_{i,u',\I'}$ where $\I' = \visit(hvu') = J \cup \visit(gvu')$. Since $\tree^{\infty}_{j,w,J}$ is part of $\forest$, it is resistant to constrained external deviations. 
    \begin{itemize}
        \item If $\gain_i(\outcome{\sigma'_{i\restriction v}, \sigma_{i \restriction hv}}{u'}) = 1$, by definition of being resistant to constrained external deviation, we have that $\gain_i(\outcome{\sigma_{\restriction h}}{v}) = 1$. In particular, we obtain that  $\gain_{i\restriction h} (\outcome{\sigma_{\restriction h}}{v}) = 1$ and $ \gain_{i \restriction h}(\outcome{\sigma'_i, \sigma_{-i \restriction h}}{v}) = \gain_{i \restriction h}(v\outcome{\sigma'_{i\restriction v}, \sigma_{-i \restriction hv}}{u'}) = \gain_{i}(\outcome{\sigma'_{i\restriction v}, \sigma_{-i \restriction hv}}{u'}) =1$.

        \item If $\gain_i(\outcome{\sigma'_{i\restriction v}, \sigma_{i \restriction hv}}{u'}) = 0$, $ \gain_{i \restriction h}(\outcome{\sigma'_i, \sigma_{-i \restriction h}}{v}) = \gain_{i \restriction h}(v\outcome{\sigma'_{i\restriction v}, \sigma_{-i \restriction hv}}{u'}) = \gain_{i}(\outcome{\sigma'_{i\restriction v}, \sigma_{-i \restriction hv}}{u'}) = 0$ since $i \not \in \visit(hv)$. Thus, in particular, as gains are either $0$ or $1$, $\gain_{i \restriction h}(\outcome{\sigma_{\restriction h}}{v}) \geq \gain_{i\restriction h}(\outcome{\sigma'_{i}, \sigma_{-i \restriction h}}{v})$.
    \end{itemize}
    As in the previous case, we can conclude that $\sigma'_i$ is not a profitable deviation in $(\game_{\restriction h},v)$.
    \end{itemize}

    We have proved that $\multiStrat$ is a permissive SPE in $\initGame$. Let us prove that $\multiStrat$ is strongly winning w.r.t.~$\winningPlayers$. Let $\sigma$ be a strategy such that $\sigma \lesssim \multiStrat$. Since $\multiHistory{\multiStrat}{\initVertex} = \tree^* = \tree_{0,\initVertex,\initVisit}$ and $\outcome{\sigma}{\initVertex} \in \tree^\infty_{0,\initVertex,\initVisit}$, we have by hypothesis that for all $i \in \winningPlayers$, $\gain_i(\outcome{\sigma}{\initVertex}) = 1$. The same kind of argument holds for the weakly winning case.

    Let us now consider the penalties. Let $i \in \playerSet$,
    \begin{align*}
        \mainPenalty_i(\multiStrat,\initVertex) &= \sup\{\penal{i}{\multiStrat}{\rho} \mid \rho \in \multiOutcome{\multiStrat}{\initVertex}\}
        = \sup\{\penalAux_i(\rho) \mid \rho \in \tree^\infty_{0,\initVertex,\initVisit} \}\\
        &= \penalAux_i(\tree_{0,\initVertex,\initVisit})
        = \mainPenalty_i(\forest) \leq m_i
    \end{align*}
    and, since for all $\tree_{i,v,\I} \in \forest$, for all $j \in \playerSet$ and for all $gu \in \tree_{i,v,\I}$, we have that $\penalAux_j(\tree_{i,v,\I \restriction gu}) \leq \penalAux_j(\tree_{i,v,\I})$. Moreover, 
    \begin{align*}
        \retaliationPenalty_i(\multiStrat,\initVertex) &= \sup_{ hv \in \hist(\initVertex) \setminus \multiHistory{\multiStrat}{\initVertex}} \sup \{ \penalAux_i(\rho) \mid \rho \in \multiOutcome{\multiStrat_{\restriction h}}{v} \}\\
        &= \sup_{\substack{ \multiHistory{\multiStrat_{\restriction h}}{v} \\ hv \in \hist(\initVertex) \setminus \multiHistory{\multiStrat}{\initVertex} }} \penalAux_i(
        {\multiHistory{\multiStrat_{\restriction h}}{v}}) \\ 
        &=\sup_{\substack{ \tree_{j,u,\J} \in \forest \\ (j,u,\J) \in \Out }}  
        \penalAux_i(\tree_{j,u,\J})
        = \retaliationPenalty_i(\forest) \tag*{\qedhere}
    \end{align*}
\end{proof}

\section{Proofs of Section~\ref{section:comput}: \nameref{section:comput}}

\bound*

\begin{proof}
    Let $\tree$ be a tree $D$-resistant to internal deviations, and $(\gamma,D)$-resistant. 
    The proof consists in adding information in the nodes of the tree, to be able to safely cut some branches in order to keep the desired guarantees while reducing the height of the symbolic tree. Besides the history of play, we thus start by labeling each node of $\tree$ (and we will keep such a labeling in all the trees we build after) by a triple $(I, (m_i)_{i\in D}, (p_i)_{i\in N'})$ where
    \begin{itemize}
    \item $N\setminus D\subseteq I \subseteq N$ is the set of players that have already won so far; 
    \item $m_i\in \{\forall, \none, ?\}$ represents what player~$i$ \textbf{m}ust guarantee in the future: $\forall$ if all subsequent plays must be winning for him (and player~$i$ has not already won so far, which is why we restrict $i$ to be in $D$, otherwise player $i$ has won already at the beginning of the tree), $\none$ if all must be losing for him, and $?$ if there are no longer or not yet any constraints; 
    \item $p_i\in \{0, 1, \ldots, P_i\}$ represents the current \textbf{p}enalty of player~$i$.
    \end{itemize}
    Notice that the signification of $\forall$ directly implies that we will never have $m_i = \forall$ and a subset $I$ containing $i$ in the same label.
    
    We label the root of the tree (that contains a vertex $v$) with a triple $(I, (m_i)_{i\in D}, (p_i)_{i\in N'})$ in such a way that $I = N\setminus D  \cup\{ i \in D\mid v\in F_i\}$, for each $i \in D$, $m_i = {?}$ if $v\in F_i$ (and $m_i$ is not constrained yet otherwise), and for each $i\in N'$, $p_i=0$. For a player that must be strongly winning (as it will be the case at some point of this proof) but $v\notin F_i$, we let $m_i = \forall$. For the other part of the proof, where we will have weakly winning constraints, we let $m_i={?}$ for all $i\in D$.

    Knowing the current label of a node, whose history ends in a vertex $v$ owned by player~$i$,
    the label $(I', (m'_i)_{i\in D}, (p'_i)_{i\in N'})$ of a successor $v'$ can be obtained as follows: 
    \begin{enumerate}[(a)]
    \item the set of winners is updated to $I' = I\cup\{j\mid v'\in F_j\}$;
    \item values $m'_j$ are set as follows: 
    \begin{itemize}
        \item if $j\in I'$, we let $m'_j={?}$. If $m_j$ was $\forall$ in the parent node, this releases the guarantees that player~$j$ had to fulfill.
        %, forbidding this update if previously we add $m_j=\none$ (this never happens in $\tree$ by $\gamma$-consistency and $D$-resistance to internal deviations);  
        Notice that since $\tree$ is $D$-resistant to internal deviations and $(\gamma,D)$-resistant, such an update cannot happen if $m_j = \none$ for the parent node;
       
        \item otherwise,
        
        \begin{itemize} \item if $j \neq i$, we keep $m'_j=m_j$ in the successor;
        
        \item otherwise, \begin{itemize}
            \item if $m_i=\forall$ or $\none$, we keep $m'_i=m_i$ in the successor; 
            \item if $m_i = {?}$ and there is $u'\in \successor(v)$ that is not a successor of the parent node $v$ in the tree with $\gamma_{i, u', I\cup\{j\mid u'\in F_j\}}=1$, we let $m'_i = \forall$;
            \item otherwise if $m_i = {?}$, and there are at least two successors, then we let $m'_i=\none$ if none of the plays of the subtree of $v$ is winning for $i$, or $m'_i=\forall$ if all the plays of the subtree are winning for $i$ (we are necessarily in one of the two cases since $\tree$ is $D$-resistant to internal deviations); 
            \item otherwise, we let $m'_i={?}$ in the successor;
        \end{itemize}  
        \end{itemize}
    \end{itemize}
    \item the new penalties are obtained from the previous one, by letting $p'_i$ be the addition of $p_i$ and the number of missing successors in $v$, while $p'_j=p_j$ for $j\neq i$.
    \end{enumerate}
    The way each component of the labels can be updated are restricted: the set of winners $I$ can only increase for the inclusion; values $m_j$ can switch once from $?$ to $\none$ or $\forall$, and from $\forall$ to $?$ for the rest of the whole branch; values $p_j$ are non-decreasing. We moreover require that a labeled tree fulfills the following validity condition: 
    \begin{enumerate}[(d)]
        \item every branch ultimately ends with a label where all values $m_i$ are $\none$ or $?$.
    \end{enumerate}

    If we obtain a labeled tree (or a labeled symbolic tree) fulfilling the rules (a), (b), (c), and (d), then the tree is $D$-resistant to internal deviations, $(\gamma,D)$-resistant, and for all $i\in N'$, the penalty of $i$ is at most $P_i$.\footnote{The reader may have recognized rules (a), (b), and (c) as a way to describe the transition relation of a top-down tree automaton, and rule (d) as an acceptance condition that every branch of the tree has to satisfy.}

    \medskip 
    The second step of the proof is to extract from the infinite labeled tree $\tree$ a finite symbolic tree fulfilling rules (a), (b), (c), and (d), and thus the desired requirements but the one on its height. We modify it afterwards so that it also fulfills the requirement on its height. To do so, we first suppose that the tree $\tree$ is strongly winning for the subset $\winningPlayers$ of players (remember that this implies that we set $m_i=\forall$ for all players $i\in \winningPlayers$ in the label of the root, except if the root vertex belongs to $F_i$). The symbolic tree will contain three parts depicted in Figure~\ref{fig:symbolic-tree}: a core taking care of the strongly winning part (which will thus ensure that the symbolic tree is also strongly winning for $\winningPlayers$); an expanded core to keep the $D$-resistance to internal deviations, the $(\gamma,D)$-resistance and the bounded penalty; a closure of branches to define the mapping $f$ sending leaves of the symbolic tree to ancestor nodes outside of the core and expanded core.

    \begin{figure}[tbp]
        \centering
        \includegraphics[height=3.5cm]{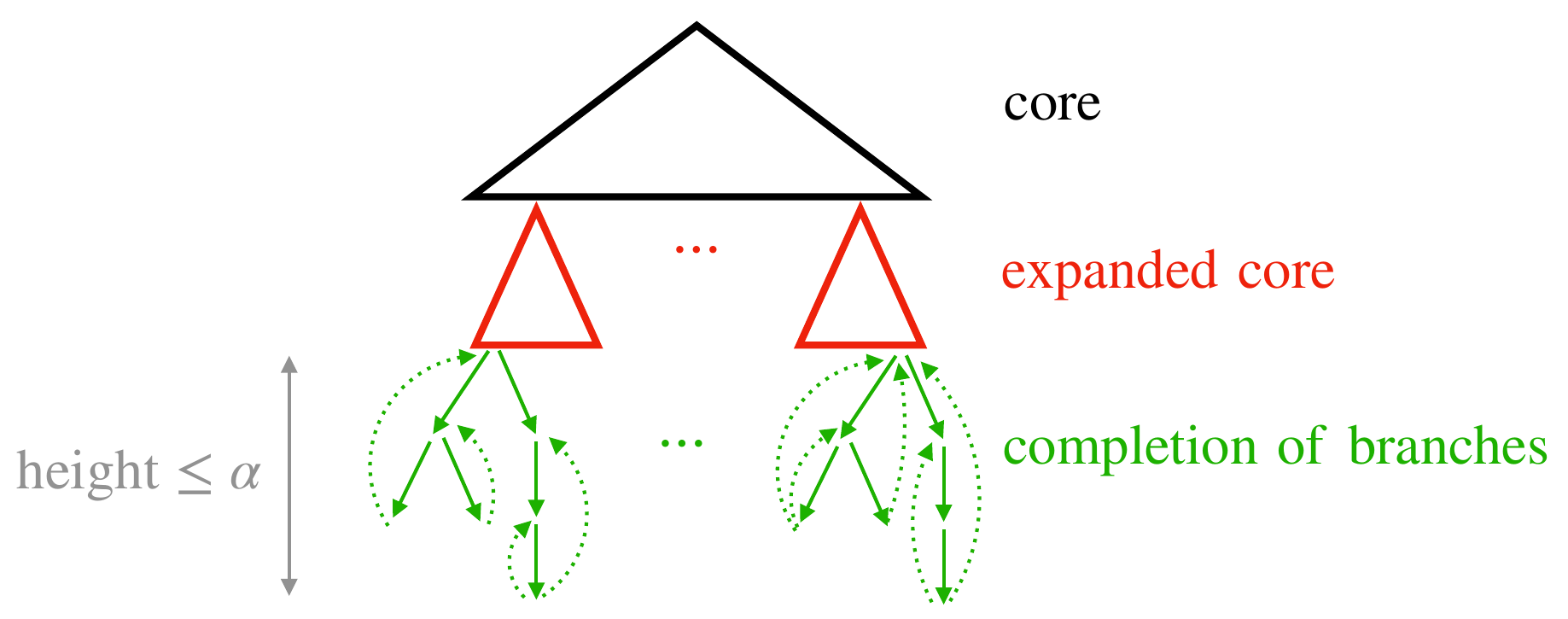}
        \caption{Construction of the symbolic tree}
        \label{fig:symbolic-tree}
    \end{figure}

    Since all players of $\winningPlayers$ match their reachability objectives in all branches, every play in $\infTree$ contains a first vertex where all players of $\winningPlayers\cap D$ have seen one of their target vertices. 
    Vertices not lower that these last targets form a finite tree that we call the \emph{core} of $\tree$. In case $\winningPlayers=\emptyset$, the core is reduced to its root. 

    We then need to complete the core to obtain a symbolic tree. This is done by considering the branches of $\tree$, alternating two steps: 
    \begin{itemize}
        \item a new player (in particular not in $\winningPlayers$) may need to win because of constraints $\forall$ in the labels: the expansion of the core because of this reason is performed as before by adding a subtree in which all branches have reached a target vertex of this player, and we call \emph{expanded core} the added subtrees;
        \item other branches must be completed until we reach a node $n$ where all successor nodes are labeled by a history ending in a vertex $v$, and by a triple label such that one of its ancestor outside of the expanded core is labeled by a history $h$ ending in the same vertex $v$, and by the same triple label. The mapping $f$ sends node $n$, which becomes a leaf in the symbolic tree, to all such $h$. We call \emph{completion of branches} these last sections of branches outside of the expanded core. Successors of (what become a) leaf of the symbolic tree could be several ancestors of this leaf, as depicted in the green part of Figure~\ref{fig:symbolic-tree} with the dotted back arrows. 
     \end{itemize}
    Notice that during the completion of branches, some new constraints $\forall$ may be added, which would then requires to return to the first step of expansion. But since this is for new players each time (and there is a finite number of players), this iterative expansion process will stop after finitely many steps. 
    
    Since there are only a finite number of labels, and the three parts of the labels cannot freely change from a value to another one, the completion of branches leads to a length bounded by $\alpha = |V| \times |D| \times 2|D| \times \sum_{i\in N'} P_i$ (in the worst case scenario, at each occurrence of a new vertex of $V$, at least one part of the labels has changed): the $|D|$ part is because the subsets $I$ all contain $N\setminus D$ and thus can only increase $|D|$ times, the $2|D|$ part is because of the switch in the $m_i$ label that can go from $?$ to $\forall$ and then to $?$ until the end.
    
    We thus have built a finite symbolic tree $\stree_1$. The labels are still consistent with the ones of $\tree$ which ensures that $\stree_1$ fulfills properties (a), (b), (c), and (d) and thus: 
    \begin{enumerate}
        \item $\widetilde \stree_1$ is $D$-resistant to internal deviations;
        \item in $\widetilde \stree_1$, each player $i\in N'$ has a penalty at most $P_i$;
        \item $\widetilde \stree_1$ is $(\gamma,D)$-resistant.
    \end{enumerate}
    Moreover, by the core of the symbolic tree, we know that $\widetilde \stree_1$ is strongly winning w.r.t.~$\winningPlayers$. 

    We now explain how to reduce the height of this symbolic tree, by reducing the height of the core and expanded core, while maintaining the rules (a), (b), (c), and (d), and not touching the completion of branches that already have a sufficient bound on its height. The obtained symbolic tree is called $\stree$.

    If the core and expanded core contains two histories ending in the same vertex $v$ with the same triple label (in particular no players see his objective for the first time in-between the two occurrences), we can modify $\stree_1$ by replacing the subtree in the first occurrence of $v$ by the subtree in the second occurrence of $v$. 
    This may remove some plays in $\widetilde\stree^\infty$ but does not change the set of winners, and does not increase the penalty of plays that remain: indeed the labels remain consistent which proves that the above properties are maintained. 
    We can continue doing this simplification step as long as possible.
    Then, if we can no longer apply the simplification, we are sure that the length of the branches in the updated expanded core is bounded by $\alpha$ (with the same parameter $\alpha$ as above). 
    We obtain a symbolic tree $\stree$ that fulfills the desired properties. 
    By adding the length of the completion of branches, $\stree$ has height at most $2\alpha$, that is  polynomial in the game and the maximal penalty $P_i$.

\medskip 
    Suppose then that $\tree$ is weakly winning w.r.t.~$\winningPlayers$, and let $\rho$ be a branch where players in $\winningPlayers$ all win. We consider the same construction as before but letting the core be the shortest prefix of $\rho$ where all players of $\winningPlayers$ have visited their target. We build the expanded core and the rest of the branches of length at most $\alpha$ as before. We then start by reducing the size of the core independently of the rest, to obtain a new core of length bounded by~$\alpha$. The expanded core is then also reduced as in the case of strongly winning trees, getting a height at most $\alpha$. In total, we obtain a symbolic tree of height at most $3\alpha$, once again polynomial in the game and the maximal penalty (when encoded in unary).
\end{proof}

\end{document}